\documentclass[pre,eqsecnum,showpacs,amssymb]{revtex4}
\usepackage{amsmath,graphicx}
\usepackage{epsfig}
\usepackage{bm}

\begin{document}

\title{Thermal Breakage and Self-Healing of a Polymer Chain under Tensile
Stress}

\author{A. Ghosh$^1$, D. I. Dimitrov$^{2}$, V. G. Rostiashvili$^1$, A.
Milchev$^{1,3}$, and T.A. Vilgis$^1$}
\affiliation{$^1$ Max Planck Institute for Polymer Research, 10 Ackermannweg,
55128 Mainz, Germany\\
$^2$ Inorganic Chemistry and Physical Chemistry
Department, University of Food Technology, Maritza Blvd. 26, 4000
Plovdiv, Bulgaria,\\
$^3$ Institute for Physical Chemistry, Bulgarian Academy of Sciences, 1113
Sofia, Bulgaria}

\begin{abstract}
We consider the thermal breakage of a tethered polymer chain of discrete
segments coupled by Morse potentials under constant tensile stress. The chain
dynamics at the onset of fracture is studied analytically by Kramers-Langer
multidimensional theory and by extensive Molecular Dynamics simulations in $1D$-
and $3D$-space. Comparison with simulation data in one- and three dimensions
demonstrates that the Kramers-Langer theory provides good qualitative
description of the process of bond-scission as caused by a {\em collective} 
unstable mode. We derive distributions of the probability for scission over the
successive bonds along the chain which reveal the influence of chain ends on
rupture in good agreement with theory. The breakage time distribution of an
individual bond is found to follow an exponential law as predicted by theory.
Special attention is focused on the recombination (self-healing) of broken
bonds. Theoretically derived expressions for the recombination time and distance
distributions comply with MD observations and indicate that the energy barrier
position crossing is not a good criterion for true rupture. It is shown that the
fraction of self-healing bonds increases with rising temperature and friction.

\end{abstract}


\maketitle

\section{Introduction}
A great variety of problems both in material and basic science rely on the
fundamental understanding of the intramolecular dynamics and kinetics of
fragmentation (bond rupture) of linear macromolecular subject to a tensile
force. Typical examples comprise material failure under stress
\cite{Kausch,Crist}, polymer rupture \cite{Garnier,Gaub,Saitta,Maroja,Rohrig},
adhesion \cite{Gersappe}, friction \cite{Klafter}, mechanochemistry
\cite{Aktah,Beyer}, and biological applications of dynamical force microscopy
\cite{Harris,Pereverzev}. In particular, the problem of polymer rupture as a
kinetic process has a longstanding history and dates back to the  publications
of Bueche \cite{Bueche} and Zhurkov {\it et al.} \cite{Zhurkov}. In these papers
the breaking of intramolecular bonds is treated as a thermally activated process
and is described by Arrhenius's formula for the rate of bond scission.

In recent years investigations have been complemented by ample application of
computer experiments. Comprehensive Molecular Dynamic (MD) simulations of $1D$
chain fragmentation (at constant strain) have been carried out, whereby harmonic
\cite{Taylor,Lee}, Morse \cite{Stember,Bolton,Sebastian}, or Lennard-Jones
\cite{Oliveira_1,Oliveira_2,Oliveira_3,Oliveira_4} chain models have been used.
One of the principal questions to be answered is : ``How long will it take for
this system to break?''.  A theoretical interpretation of MD-results
\cite{Oliveira_3,Oliveira_4}, based on an effectively one-particle model, has
been suggested in terms of Kramers rate theory \cite{Kramers}. Thus, the
activation energy of the one-particle model $E_{b}$ has been found to be close
to the barrier height observed in the simulations, whereas the measured
frequency of bond scission appears more than two orders of magnitudes
smaller then the corresponding frequency, predicted by Kramers theory.
The authors interpret this controversial result as a manifestation of the
existence of collective modes which are missing in the one-particle Kramers
theory.

It has also been noticed that a truly irreversible break may occur, if bonds
are stretched to lengths, considerably larger than the one corresponding to the
barrier position. This indicates the possibility for bond recombination whereby
the chain integrity is restored  with some finite probability.

The problem of polymer fragmentation has also been studied
theoretically \cite{Sebastian} for the case of constant stress when a tethered
chain of segments is subjected to a pulling force at the free chain end. The
consideration has been based on a multidimensional version of the transition
state theory (TST). Friction is then taken into account by coupling the polymer
to a set of harmonic oscillators, simulating thus the presence of a thermostat.
A comparison of the calculated breaking rate with the corresponding MD
observation \cite{Oliveira_1} shows again that the theoretical rate is about 250
times larger.  Also the role of the bond healing process has been discussed 
which helps to improve to some extent the agreement between theory and
simulation. Nevertheless, despite the multidimensional nature of TST, it does
not take into account properly the collective unstable mode development, which
leads, in our opinion, to the essential overestimation of the breaking rate.

The collectivity effect has been recently treated \cite{Sain} for constant
strain and periodic boundary conditions (a ring polymer) on the basis of the
multidimensional Kramers approach \cite{Langer,Hanngi}. Within this approach the
development of a collective unstable mode and the effect of dissipation can be
described consistently. It has been shown that in this case the effective
break frequency is of the same order of magnitude as the one observed in the
simulation.

In the present work we develop this approach further for the case of a tethered
Morse chain, consisting of $N$ segments and subjected to a constant tensile
force $f$ applied at its free end. We derive analytic expressions for the
scission rate of the bonds, its distribution along the polymer chain, and its
variation with changing temperature and dissipation. For comparison with
computer experiment, we also perform extensive MD simulations in both one- $1D$,
and three dimensions, $3D$, and witness significant differences in the
fragmentation behavior of the chain, despite the observed good agreement
between theoretical predictions and simulation data. A major objective of the
current study is the elucidation of the problem of bond recombination
(self-healing) which has found little attention in literature so far. To this
end we derive analytic expressions for the life times and extension distances
of healing bonds and compare them to our MD results.

The paper is organized as follows. In Section \ref{Kramers_Langer} we give a
sketch of the multidimensional Kramers-Langer escape  theory
\cite{Langer,Hanngi}. We also outline the problem of multiple points of exit
from potential well \cite{Matkowsky,Gardiner} which is necessary for treating
bond rupture with respect to the consecutive number of each bond along the
chain. In Section \ref{Model_description} we present our model of
one-dimensional Morse string of beads and consider the  eigenvalue problem in
the vicinity of the metastable minimum of the effective potential and at the
barrier (saddle point) which is needed for the description of the  unstable
collective mode. Section \ref{MD_results} gives briefly details of our MD
simulation and presents the main numeric results as well as their interpretation
in the light of our theoretical approach. In Section \ref{Self_Healing} the
healing process is discussed in terms of distributions of healing times and bond
extensions. We give also the theoretical interpretation of this process based on
the solution of the Kramers equation \cite{Risken} using an inverted harmonic
potential for representation of the  barrier. We conclude in Section
\ref{Summary} and outline some future developments.

\section{Kramers-Langer multidimensional escape theory}
\label{Kramers_Langer}

\subsection{Rate of escape}\label{Kramers_Rate}

The calculation of the rate of escape is based on the $N$-dimensional (in the
total phase space $x_i = q_i$, $x_{i+N} = p_i$, $i\leq N$ \cite{Goldstein})
Fokker-Planck equation \cite{Risken}
\begin{eqnarray}
\dfrac{\partial P({\bf x}, t)}{\partial t} =  \sum_{i, j=1}^{2 N}
\:\dfrac{\partial }{\partial x_i} \:
M_{i j} \left[\dfrac{\partial H}{\partial x_j} \:  P({\bf x}, t)  + T
\dfrac{\partial}{\partial  x_j} \:  P({\bf x}, t)
\label{F_P}\right]
\end{eqnarray}
for the probability distribution function $P({\bf x}, t)\equiv P(\{x_i\}, t)$.
In eq. (\ref{F_P}) the Hamiltonian has a general form $\sum_{i=1}^{N}
\: p_i^2/2 m + V(\{q_i\})$. The $2 N\times 2 N$-matrix $M_{i j} = \Gamma_{i j}
- A_{i j}$ , where $\Gamma_{i j}$ is the matrix of Onsager coefficients and
$2N\times 2N$ skew-symmetric  matrix \cite{Goldstein}
\begin{eqnarray}
 \overrightarrow{A} = \left(\begin{array}{ccc}
                       \overrightarrow{0}& \overrightarrow{1}\\
-\overrightarrow{1}& \overrightarrow{0}
                      \end{array}\right)
\end{eqnarray}
where $\overrightarrow{1}$ is the $N\times N$ unit matrix
 and $\overrightarrow{0}$ is the $N\times N$ zero matrix.
The eq.(\ref{F_P}) can be seen as a continuity equation, $\partial P
/\partial t = -
\sum_{i=1}^{2 N} \: \partial / \partial x_i \: J_i$,   where the probability
current is given by $ J_i =-\sum_{j=1}^{2 N} M_{i j}\left(\partial
H/\partial x_j P({\bf x}, t)  + T \partial P({\bf x}, t)/\partial x_j \right)$.

It is assumed \cite{Langer} that there is a metastable state $\{ x_i^{A}\}$
which is separated with a barrier from another stable state. The coordinates at
the {\it saddle point} which separates these two states is denoted by
$\{x_i^{S}\}$. The escape from the metastable minima is a comparatively rare
event, so that one can treat the process close to $\{x_i^{S}\}$ as a stationary
one, i.e. $\sum_{j=1}^{2 N} \partial J_j/\partial x_j = 0$.  Thus one gets
\begin{eqnarray}
 \sum_{i, j=1}^{2 N} \: \dfrac{\partial }{\partial x_i}\: M_{i
j}\:\left[\sum_{k} {\rm E}_{j k}^S (x_k - x_k^{S}) + T
\dfrac{\partial}{\partial x_j} \:  \right] P({\bf x}) = 0
\label{Stationary}
\end{eqnarray}
where also the harmonic approximation around $\{x_i^{S}\}$ has been used, i.e.,
the Hamiltonian reads
\begin{eqnarray}
 H(\{x_i\}) = E^S + \dfrac{1}{2} \sum_{j, k =1}^{2 N} {\rm E}_{j k}^S (x_j -
x_j^{S}) (x_k - x_k^{S})
\label{Harmonic}
\end{eqnarray}
where $E^S = H(\{x_i^{S}\})$ and the Hessian matrix $E_{i j}^S = \partial^2
H/\partial x_i \partial x_j$ at $\{x_i\} = \{x_i^{S}\}$.

One should impose the following boundary conditions. Near the metastable state
$\{x_i^{A}\}$  the distribution function $P({\bf x})$ is the equilibrium
one, i.e.
\begin{eqnarray}
 P({\bf x}) = P_{\rm eq}({\bf x}) = Z_A^{-1} \: \exp (- \beta H) \qquad
\mbox{at} \qquad {\bf x} \simeq \{x_i^A\}
\label{BC_1}
\end{eqnarray}
On the other hand, all states around the stable minimum (which is far beyond the
top of the barrier!) are removed by a sink, i.e.,
\begin{eqnarray}
  P({\bf x}) \simeq 0 \qquad \mbox{at} \quad  \{x_i\} \quad  \mbox{far
beyond} \quad \{x_i^S\}
\label{BC_2}
\end{eqnarray}
With a transformation to new coordinates $\xi_n  = \sum_{i=1}^{2N} D_{n
i} (x_i - x_i^{S})$, which are principal-axis coordinates for the Hamiltonian
given by eq.(\ref{Harmonic}), one obtains
\begin{eqnarray}
 H({\bm \xi}) = E^{S} + \dfrac{1}{2} \: \sum_{n = 1}^{2 N} \: \lambda_n \:
\xi_n^2
+ \dots
\label{Hamiltonian_Normal}
\end{eqnarray}
where the vector ${\bm \xi} \equiv \{\xi_n\}$. The matrix $D_{n i}$  is
orthogonal one, i.e. $D^T =
D^{-1}$ and $\{\lambda_n\}$ are the eigenvalues of the matrix ${\rm E}_{i
j}^S$.  Since $\{x_i^S\}$ is a saddle point one of the $\lambda$'s, say
$\lambda_1$, is negative. The standard trick would be to look for the solution
in the form $P({\bm \xi}) = W({\bm \xi})  P_{\rm eq}({\bm \xi})$ , where
$W({\bm \xi})$ is a new function. Thus in the $\xi$-coordinate system the
steady - state Fokker Planck equation  for $W({\bm \xi})$ takes the form
\begin{eqnarray}
 \sum_{n, k=1}^{2 N} \:\left({\widetilde{\Gamma}}_{n k}\:\dfrac{\partial^2
W}{\partial \xi_n \partial \xi_k} - \beta \lambda_n \xi_n \: {\widetilde{M}}_{n
k} \dfrac{\partial W}{\partial \xi_k}\right) = 0
\label{FP_Harmonic}
\end{eqnarray}
where ${\widetilde{M}}_{n k} = \sum_{i,j} \: D_{n i} M_{i j} D_{k j} =
{\widetilde{\Gamma}}_{n k} - {\widetilde{A}}_{n k}$.

In the same manner as for the one-dimensional Kramers problem \cite{Kramers} one
could claim that the function $W({\bm \xi})$ depends only on a linear
combination of all $\xi_n$, i.e. $W(\{\xi_k\}) = F(u)$, where $u =
\mathop{{\sum}'}_{n} \: U_n \: \xi_n$ . The prime in this expression  indicates
that we omit all $n$'s for which $\lambda_n=0$. The resulting equation reads:
$\sum_{n,k} \left[({\widetilde{\Gamma}}_{n k}\: U_n \; U_k) \: d^2 F/d u^2 -
\beta(\lambda_n \: \xi_n  {\widetilde{M}}_{n k}\: U_k)  d F/d u \right] = 0$. As
in the one-dimensional Kramers problem \cite{Kramers} the coefficient in front
of $d F/d u$ is a linear function of $u$, i.e.
$\sum_{n,k} \lambda_n \: \xi_n  {\widetilde{M}}_{n
k}\: U_k) = \kappa u = \kappa  \mathop{{\sum}'}_{n} \: U_n \: \xi_n$. As a
result
\begin{eqnarray}
 \lambda_n \: \sum_{k} {\widetilde{M}}_{n
k}\: U_k = \kappa \: U_n
\label{Eigenproblem}
\end{eqnarray}
i.e., the coefficients $U_n$ are solutions of the eigenvalue problem eq.
(\ref{Eigenproblem}). Thus,
\begin{eqnarray}
 T \: \lambda_{+} \: \dfrac{d^2 F(u)}{d u^2} - u \dfrac{d F(u)}{d u} = 0
\label{F}
\end{eqnarray}
where
\begin{eqnarray}
\lambda_{+}  = \dfrac{1}{\kappa} \: \sum_{n,k} \: U_n \: {\widetilde{\Gamma}}_{n
k}\: U_k
\label{Lambda}
\end{eqnarray}

There is a simple physical interpretation of the eigenvalue problem given by eq.
(\ref{Eigenproblem}) \cite{Langer}. Namely, the equation of motion for the
average value, $\left\langle \xi_n(t)\right\rangle = \int \prod_{j=1}^{2N} \xi_n
P(\{\xi_j\}, t)$ can be written as
\begin{eqnarray}
 \dfrac{\partial }{\partial t} \left\langle   \xi_n (t) \right\rangle = \int
\prod_{j=1}^{2N} \: d \xi_j \: \widetilde{J} (\{\xi\}, t) = -
\dfrac{T}{Z_A} \sum_{k} \: {\widetilde{M}}_{n k}\:\int \prod_{j=1}^{2N} \: d
\xi_j \: \dfrac{\partial W}{\partial \xi_k} \: {\rm e}^{-\beta H}
\end{eqnarray}
With integration by parts and the harmonic approximation, eq.
(\ref{Hamiltonian_Normal}), one arrives at the following expression
\begin{eqnarray}
 \dfrac{\partial }{\partial t} \left\langle   \xi_n (t) \right\rangle = -
\sum_{k} \:{\widetilde{M}}_{n k}\:\lambda_k \: \left\langle   \xi_k (t)
\right\rangle
\label{Average}
\end{eqnarray}
The unstable solution of this equation (which describes the decay of a
metastable state) is given by a negative eigenvalue $\kappa$, namely
\begin{eqnarray}
 \left\langle   \xi_n (t) \right\rangle = X_n \: {\rm e}^{- \kappa t}
\label{Solution}
\end{eqnarray}
Substitution of eq.(\ref{Solution}) in eq. (\ref{Average}) leads to $\sum_{k}
\: {\widetilde{M}}_{n k}\: \lambda_k \: X_k = \kappa X_n$ or
\begin{eqnarray}
 \lambda_n \:\sum_{k} {\widetilde{M}}_{n k}\: \lambda_k \: X_k = \kappa
\lambda_n X_n
\label{U}
\end{eqnarray}
This equation is identical to eigenvalue problem, eq. (\ref{Eigenproblem}),
provided that $U_n = \lambda_n X_n$. There is a negative eigenvalue $\kappa$
and as a result a negative $\lambda_{+}$ (see eq.(\ref{Lambda})) which
corresponds to the unstable mode. This clear physical interpretation justifies
 the linear combination  ansatz which has been used upon the derivation of eq.
(\ref{F}). We will show in Sec. IV devoted to  MD-simulation results that
the law given by eq. (\ref{Solution}) actually holds for the breaking bonds.

With the negative eigenvalue, $\lambda_{+} < 0$, the solution of eq.(\ref{F})
takes the form
\begin{eqnarray}
 F(u) = \dfrac{1}{\sqrt{2 \pi |\lambda_{+}| T}} \: \int_{u}^{\infty} \: d
z\:  \exp \left( - \dfrac{z^2}{2 |\lambda_{+}| T}\right)
\label{Solution_F}
\end{eqnarray}
In eq.(\ref{Solution_F}) we take into account the boundary conditions,
eqs.(\ref{BC_1}) and (\ref{BC_2}) which require $F (u \rightarrow -
\infty) = 1$ (at the metastable well) and $F (u \rightarrow \infty) = 0$
(around the stable minimum).

Consider the rate of the metastable state decay. The main quantity which
should be used for this purpose is the the probability current
\begin{eqnarray}
 {\widetilde{J}}_n = - \dfrac{T}{Z_A} \: \sum_{k} {\widetilde{M}}_{n k}\: U_{k}
\: \dfrac{d F (u)}{d u} \: {\rm e}^{- \beta H}
\label{Current_2}
\end{eqnarray}
To obtain the total probability flux over the barrier one should integrate the
 current, eq. (\ref{Current_2}), over the hypersurface $u =
\mathop{\sum'}_{n} U_n \xi_n = 0$ containing the saddle point. The resulting
flux reads
\begin{eqnarray}
 J = \int \prod_{i = 1}^{2 N} \: d \xi_i \: \delta (u) \: \sum_{n=1}^{2N} \: U_n
{\widetilde{J}}_n (\{\xi_i\})
\label{Flux}
\end{eqnarray}
The calculation of this integral is given in Appendix of ref. \cite{Hanngi}. The
calculation yields
\begin{eqnarray}
 J = \dfrac{|\kappa|}{2 \pi} \: \left(\dfrac{2 \pi T}{|\lambda_1|}\right)^{1/2}
\: \prod_{n=2}^{2 N} \left(\dfrac{2 \pi T}{\lambda_n}\right)^{1/2} \:
\dfrac{{\rm e}^{-\beta E^S}}{Z_A}
\label{Flax_2}
\end{eqnarray}

In order to obtain the rate constant one must divide the flux over the
popularion $n_A$ in the metastable well
\begin{eqnarray}
 n_A = \int\limits_{A_{well}}  \: \prod_{i=1}^{2 N} \: d \xi_i P_{\rm eq}
(\{\xi_i\}) = \dfrac{{\rm e}^{- \beta E^A}}{Z_A} \: \prod_{i=1}^{2 N}
\:\left(\dfrac{2 \pi
T}{\lambda_n^a}\right)^{1/2}
\label{Population}
\end{eqnarray}
where we have used the harmonic approximation near the metastable state
$\{x_i^{A}\}$, i.e. $H = E^{A} + \sum_{n=1}^{2N} \lambda_n^{a} \xi_n^{2}/2$.
Combining eq.(\ref{Flax_2}) and eq.(\ref{Population}), one  finds for the rate
constant, $k = J/n_A$, the following result \cite{Langer}
\begin{eqnarray}
  k = \dfrac{|\kappa|}{2 \pi} \:\:\left[\dfrac{{\rm det}({\bf E}^A/2 \pi
T)}{|{\rm
det}({\bf E}^{S}/2 \pi T)|} \right]^{1/2} \: {\rm e}^{- \beta E_b}
\label{Rate_Constant_3}
\end{eqnarray}
where the activation barrier $E_b = E^S - E^A$. We recall that $\lambda_n$ are
the eigenvalues (without zero-modes) of the Hessian matrix ${\bf E}^S$ at the
saddle point ${\bf x}^S$ whereas $\lambda_n^a$ are the the eigenvalues of the
Hessian matrix ${\bf E}^A$ at the metastable point ${\bf x}^A$.

\subsection{The eigenvalue problem}

Consider in more detail the eigenvalue problem given by eq. (\ref{U}),
i.e.$\sum_{k} {\widetilde{M}}_{n k}\: \lambda_k \: X_k = \kappa
 X_n$. In the initial $x$-space this equation reads
\begin{eqnarray}
 \sum_{k, r} M_{n k} \: E_{k r}^S \: X_r = \kappa X_n
\label{U_3}
\end{eqnarray}
where $M_{i j} = \Gamma_{i j} -  A_{i j}$ and the friction $\Gamma_{i j}$ as
well as the matrix $A_{i j}$ are given as
\begin{eqnarray}
 \Gamma_{i j} = m \gamma \left(\begin{array}{ccc}
 \begin{array}{ccc}
 0 &\cdots & 0\\
\vdots &   & \vdots\\
0 & \cdots & 0
\end{array} &   &
\begin{array}{ccc}
 0 &\cdots & 0\\
\vdots &  \ddots & \vdots\\
0 & \cdots & 0
\end{array}
\\
 \begin{array}{ccc}
 0 &\cdots & 0\\
\vdots &  \ddots & \vdots\\
0 & \cdots & 0
\end{array}&  &\begin{array}{ccc}
 1&\cdots & 0\\
\vdots &  \ddots  & \vdots\\
0 & \cdots & 1
\end{array}
 \end{array}\right) \qquad
A_{i j} = \left(\begin{array}{ccc}
 \begin{array}{ccc}
 0 &\cdots & 0\\
\vdots &   & \vdots\\
0 & \cdots & 0
\end{array} &   &
\begin{array}{ccc}
 1 &\cdots & 0\\
\vdots &  \ddots & \vdots\\
0 & \cdots & 1
\end{array}
\\
 \begin{array}{rcr}
 -1 &\cdots & 0\\
\vdots &  \ddots & \vdots\\
0 & \cdots & -1
\end{array}&  &\begin{array}{ccc}
 0&\cdots & 0\\
\vdots &    & \vdots\\
0 & \cdots & 0
\end{array}
 \end{array}\right)
\label{Gamma_and_A}
\end{eqnarray}
The $2N \times 2N$ Hessian matrix and the $2 N$-dimensional column-vector are
\begin{eqnarray}
E_{i j}^S  =
\left(\begin{array}{ccc}
  \dfrac{\partial^2 V}{\partial q_i \partial q_j}&   &
\begin{array}{ccc}
 0 &\cdots & 0\\
\vdots &   & \vdots\\
0 & \cdots & 0
\end{array}
\\
 \begin{array}{ccc}
 0 &\cdots & 0\\
\vdots &   & \vdots\\
0 & \cdots & 0
\end{array}&  &\begin{array}{ccc}
 1/m &\cdots & 0\\
\vdots &  \ddots & \vdots\\
0 & \cdots & 1/m
\end{array}
 \end{array}\right) \qquad
 X_i = \left(\begin{array}{c}
q_1 \\
q_2\\
\vdots\\
q_N\\
p_1\\
p_2\\
\vdots\\
p_N
\end{array}\right)
\label{Hessian_and_Column}
\end{eqnarray}
After substitution of eqs. (\ref{Gamma_and_A}) and
(\ref{Hessian_and_Column}) into eq. (\ref{U_3}), and exclusion of momentums
$\{p_i\}$, one arrives at the eigenvalue problem
\begin{eqnarray}
 \sum_{j=1}^N \: \underbrace{\left[V_{i j}^S - m \gamma \kappa \delta_{ij}
\right]}_{\mbox{$N\times N$ - matrix}} \: X_j = - m  \kappa^2 X_i
\label{Final_Eigenvalue}
\end{eqnarray}
where the notation $V_{i j}^S = \partial^2 V/\partial q_i
\partial q_j |_S$ is used. The corresponding characteristic equation
\cite{Lancaster}
reads
\begin{eqnarray}
 \det \left[m \kappa^2 \delta_{i j} - m \gamma \kappa \delta_{i j} + V_{i j}^S
\right] = \prod_{k=1}^{N} [m \kappa^2 - m \gamma \kappa + \lambda_k] = 0
\label{Characteristic_Equation}
\end{eqnarray}
where $\{\lambda_k\}$ is a set of eigenvalues of the Hessian $V_{i j}^S$. As
mentioned before, only one eigenfunction, say $\lambda_1$, is negative.
This eigenvalue determines then the equation for the {\it transmission
factor} $\kappa$, i.e.
\begin{eqnarray}
 \kappa^2 - \gamma \kappa - \dfrac{|\lambda_1|}{m} = 0
\end{eqnarray}
The negative solution reads
\begin{eqnarray}
 \kappa = - \sqrt{ \dfrac{\gamma^2}{4} + \dfrac{|\lambda_1|}{m}} +
\dfrac{\gamma}{4} \: < 0
\label{Kappa}
\end{eqnarray}
Eq. (\ref{Characteristic_Equation}) has been discussed first in ref.
\cite{Weidenmuller}.

\subsection{First-passage-time approach} \label{FPT_Approach}

\begin{figure}[ht]
\begin{center}
\includegraphics[scale=0.5, angle=0]{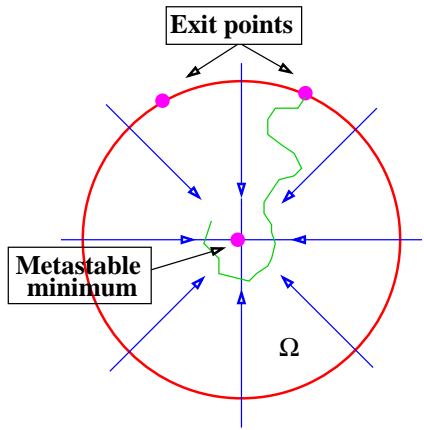}
\caption{Exit from the phase space domain $\Omega$. Vectors denote the restoring
force directions. The trajectory starts close to the metastable minimum and
exits through a particular exit (saddle) point.}
\label{Attraction_Domain}
\end{center}
\end{figure}

The method of Section \ref{Kramers_Rate} is a quite general approach to evaluate
the rate of escape from the metastable state. The alternative to the
flux-over-population method involves the concept of mean first-passage time. For
an arbitrary stochastic multi-dimensional process $ {\bf x}(t)$  the mean
first-passage time  (MFPT) $T_1 (x)$ is defined as the average time elapsed
until the process starting out at point ${\bf x}$ leaves a prescribed domain
$\Omega$ which includes the point ${\bf x}^A$ of the reactant state. $\Omega$ is
sometimes called domain of attraction of the metastable state,  and is shown in
Fig.
\ref{Attraction_Domain}

For the FPT investigation it is convenient to use
the Fokker-Planck equation (for the conditional probability $P({\bf y}, t;
{\bf x}, 0)$ of visiting the point ${\bf y} \in \Omega$ at time $t$, provided it
starts from ${\bf x} \in \Omega$ at $t=0$) which includes the so called
{\it adjoint operator} and has the form \cite{Risken}
\begin{eqnarray}
  \dfrac{\partial }{\partial t} \: P({\bf y}, t; {\bf x}, 0) = L^{\dagger} (
{\bf x}) \:
P({\bf y}, t; {\bf x}, 0)
\label{Fokker-Planck_Adjoint}
 \end{eqnarray}
 where the adjoint operator
\begin{eqnarray}
  L^{\dagger}({\bf x}) =  \sum_{i} \: K_i({x}) \: \dfrac{\partial}{\partial
x_i}
\: +
\varepsilon \: \sum_{i, j} \: D_{i j}({\bf x}) \: \dfrac{\partial^2}{\partial
x_i
\partial x_j} \:
\label{Operator_Adjoint}
 \end{eqnarray}
In the present case the drift and diffusion coefficients are respectively:
$K_i({\bf y}) = - \sum_{j} M_{i j} \partial H/\partial y_j$ and $D_{i j} =
\Gamma_{i j}$.  In eq. (\ref{Operator_Adjoint}) $\varepsilon$ is responsible for
the noise intensity (in our case $\varepsilon$ plays the role of the
temperature), so that at $\varepsilon \rightarrow 0$ the noise is weak.

The $n$-th order moments of the first-passage time, $T_n ({\bf x})$, may be
iteratively expressed in terms of the adjoint operator,
\cite{Risken}
\begin{eqnarray}
 L^{\dagger}({\bf x}) \: T_n ({\bf x}) &=& - n T_{n-1}({\bf x}) \quad, \quad
\mbox{at}
\quad {\bf x} \in \Omega \nonumber\\
 T_n ({\bf x}) &=& 0 \quad, \quad \mbox{at} \quad  {\bf x} \in \partial \Omega
\label{Iteration}
\end{eqnarray}
where the second equation simply means that the first passage time is zero,
provided the trajectory starts at the separatrix $\partial \Omega$. The
hierarchy given by eq. (\ref{Iteration})  should be supplemented by the initial
condition $T_0 ({\bf x}) = 1$ which is evident from the normalization condition
for  the FPT distribution. Therefore, the equation for the
mean first passage time (MFPT), $T_1 ({\bf x})$, reads
\begin{eqnarray}
  L^{\dagger}({\bf x}) \: T_1 ({\bf x}) &=& -1 \quad, \quad \mbox{at}
\quad  {\bf x} \in \Omega \nonumber\\
T_1 ({\bf x}) &=& 0  \quad, \quad \mbox{at} \quad  {\bf x}  \in \partial \Omega
\label{Mean}
\end{eqnarray}

On the other hand, for small noise, $\varepsilon \rightarrow 0$, a trajectory
starting within  $\Omega$ will typically first approach the attractor and stay
within its neighborhood for a long time until an occasional fluctuation drives
it to the separatrix $\partial \Omega$. Hence, MFPT  $T_1 ({\bf x})$ assumes the
same large value $\tau$ everywhere in $\Omega$ except for a thin layer along the
boundary $\partial \Omega$. Under these condition  the solution of the
hierarchical equation, eq.(\ref{Iteration}), can be obtained in  the following
form \cite{Talkner}: $T_1 = \tau , T_2 = 2 \tau^2 , T_3 = 2 \cdot 3\tau^3,
\ldots, T_n = n ! \: \tau^n$. In result the probability density of the first
passage times reads \cite{Talkner}
\begin{eqnarray}
 p(t) = \tau^{-1} \exp (- t/\tau)
\label{PDF}
\end{eqnarray}
(we recall that $\tau^{-1} \int_{0}^{\infty} t^n \exp(- t/\tau) dt = n!
\tau^n$). The validity of this PDF has been  shown in our MD-simulation (see
Sec. IV). Finally, it can be proven \cite{Talkner,Hanngi} that the
Kramers escape rate $k = (2\tau)^{-1}$.

\subsection{Distribution of exit points}

Assume that there are a number of saddle points, ${\bf b}^r$ where $r =
1,2, \dots M$, (referred to as {\em exit points})which lie on the separatrix.
We study then the exit events (see Fig. \ref{Attraction_Domain} where only two
exit points are shown). One may ask : ``What is the probability
to exit through a particular exit point ${\bf b}^S \in \partial \Omega$
irrespective of the time it takes?'' This problem has been discussed first by
Matkowsky and Schuss \cite{Matkowsky} and later by Gardiner \cite{Gardiner}.
Here we give an explicit solution which may be compared to MD-simulations.

The probability for escape through an exit point ${\bf b}^S$ if the
random trajectory starts at ${\bf x} \in \Omega$ is defined as
\begin{eqnarray}
 \pi ({\bf b}^S; {\bf x}) = \int\limits_{0}^{\infty} \: d t' \sum\limits_{i}
\: \nu_i ({\bf b}^S) J_i ({\bf b}^S, t'|{\bf x}, 0)
\label{Definition}
\end{eqnarray}
where $\nu_i ({\bf b}^S)$ is  a component of the vector normal to the separatrix
unit vector at ${\bf b}^S$ pointing out of region $\Omega$. The flux $J_i ({\bf
b}^S, t'|{\bf x}, 0)$ counts only trajectories which start at $t = 0$ in ${\bf
x} \in \Omega$ and approach the exit point ${\bf b}^S$ at the boundary
(separatrix) at time moment $t'$. The integral over time in eq.
(\ref{Definition}) implies that this probability is calculated irrespective of
the time needed for escape.

The probability $\pi ({\bf b}^S; {\bf x})$ is governed by the backward
stationary FPE, i.e. (see Sec. 5.4.2 in \cite{Gardiner})
\begin{eqnarray}
 L^{\dagger} ({\bf x}) \: \pi ({\bf b}^S; {\bf x}) = 0
\label{Governed_By}
\end{eqnarray}
The boundary conditions are: $\pi ({\bf b}^S; {\bf b}^S)=1$ and $\pi
({\bf b}^S; {\bf x}) = 0$ for any ${\bf x} \in \partial \Omega$ if ${\bf x}
\neq {\bf b}^S$, i.e.,
\begin{eqnarray}
 \pi ({\bf b}^S; {\bf x}) = \delta_s ({\bf b}^S - {\bf x}) \qquad \mbox{for}
\qquad  {\bf x} \in \partial \Omega
\label{B_C}
\end{eqnarray}

Using the same arguments as in Sec. \ref{FPT_Approach}, one can show that in the
small noise limit, $\varepsilon\rightarrow 0$,  the probability  $\pi ({\bf
b}^S; {\bf x})$ remains the same everywhere inside $\Omega$ apart from a thin
layer along the boundary $\partial \Omega$. Now we specify the general eq.
(\ref{Governed_By}) for the special case when the drift term has the form of
potential (i.e., it is derivative of the Hamiltonian). Then
eq. (\ref{Governed_By}) reads
\begin{eqnarray}
\sum\limits_{i, j}\left[ - M_{i j} \dfrac{\partial H}{\partial x_j} \:
\dfrac{\partial}{\partial x_i} + \varepsilon M_{i j} \:
\dfrac{\partial^2}{\partial x_i \partial x_j} \right]  \pi ({\bf b}^S; {\bf x})
= 0
\label{In_Coordinates}
\end{eqnarray}
Close to the saddle point the Hamiltonian can be treated in the harmonic
approximation, eq. (\ref{Harmonic}),  and the drift velocity  in eq.
(\ref{In_Coordinates}) becomes
\begin{eqnarray}
K_i = - \sum\limits_{j} \: M_{i j} \dfrac{\partial H}{\partial x_j} = -
\sum\limits_{j, k} \: M_{i j} E_{j k}^S (x_k - b_k^S)
\label{Drift_Velocity}
\end{eqnarray}

Since the solution changes only within a thin layer along the boundary (in 
direction normal to the boundary layer!), one may introduce new local
coordinates $\{z, y_p\}$ where $z$ measures the distance from ${\bf b}^S$ and
$\{y_p\}$ is the set of tangential variables measuring the orientation around
${\bf b}^S$. The coordinates $z=z({\bf x})$ and $y_p=y_p({\bf x})$ are chosen so
that
\begin{eqnarray}
 \nabla z ({\bf u}) &=& \bm{\nu}({\bf u})\nonumber\\
\bm{\nu}({\bf u})\cdot \nabla y_p({\bf u}) &=& 0\nonumber\\
z ({\bf b}^S)&=& 0
\label{Coordinates}
\end{eqnarray}
where ${\bf u} \in \partial \Omega$. The first equation in (\ref{Coordinates})
means that the coordinate $z$ changes in direction of the $\bm{\nu}$-vector. The
second equation implies that the coordinates $\{y_p\}$ are parallel to $\partial
\Omega$. In terms of new variables
\begin{eqnarray}
 \nabla_i \pi = \nu_i \dfrac{\partial \pi}{\partial z} + \sum\limits_{p}
\nabla_i y_p ({\bf x}) \dfrac{\partial \pi}{\partial y_p}
\label{Transformation_1}
\end{eqnarray}
and
\begin{eqnarray}
 \nabla_i \nabla_j  \pi &=& \nu_i \: \nu_j \: \dfrac{\partial^2 \pi }{\partial
z^2} + 2 \sum\limits_{p} \nu_i \: \nabla_j y_p ({\bf x}) \: \dfrac{\partial^2
\pi }{\partial z \partial y_p} + \sum\limits_{p, s} \:  \nabla_i y_p ({\bf x})
\nabla_j y_s ({\bf x})\: \dfrac{\partial^2 \pi }{\partial y_p \partial y_r}
\nonumber\\
&+& \nabla_i \nabla_j z({\bf x}) \: \dfrac{\partial \pi}{\partial z}  +
\sum\limits_{p} \nabla_i \nabla_j y_p({\bf x}) \: \dfrac{\partial \pi}{\partial
y_p}
\label{Transformation_2}
\end{eqnarray}
One should keep in mind that for the exit event at $\varepsilon \rightarrow 0$
only one coordinate is relevant, namely, the one traversing the saddle point in
direction of $\bm{\nu}$ vector. This is our new $z$-coordinate, which
parameterizes the displacement from the saddle point as follows
\begin{eqnarray}
 x_j &=& b_j^S + z \: \nu_j\nonumber\\
z &=& \sqrt{\varepsilon} \rho
\label{Deviation}
\end{eqnarray}
Substituting eqs.(\ref{Transformation_1})- (\ref{Deviation}) in eqs.
(\ref{In_Coordinates}) and (\ref{Drift_Velocity}), and keeping only the lowest
order in $\varepsilon$, leads to
\begin{eqnarray}
 \lambda_{+} \: \dfrac{\partial^2 \pi}{\partial \rho^2} - \rho \:
\dfrac{\partial \pi}{\partial \rho} = 0
\label{Lowest_Order}
\end{eqnarray}
where
\begin{eqnarray}
 \lambda_{+} = \dfrac{1}{\kappa} \: \sum_{i, j} \: \nu_i \: \Gamma_{i j} \:
\nu_j = \dfrac{m \gamma}{\kappa}\;.
\label{Lambda_Plus}
\end{eqnarray}
In (\ref{Lambda_Plus})one has used $\sum_{i, j} \: \nu_i \: M_{i j} \:
\nu_j = \sum_{i, j} \: \nu_i \: \Gamma_{i j} \:
\nu_j $ and taken into account $\sum_{i, j} \; \nu_i \: \Gamma_{i
j} \: \nu_j
= m \gamma$ (Appendix \ref{Calculation}) as well as
\begin{eqnarray}
 \kappa = \sum\limits_{i,j,k} \: \nu_i \:M_{i j} E_{j k}^S \: \nu_k
 \end{eqnarray}

The solution of eq.(\ref{Lowest_Order}) has the form
\begin{eqnarray}
\pi ({\bf b}^S; {\bf u}, \rho) = \delta ({\bf u} - {\bf b}^S) + [C_{\infty} -
\delta ({\bf u} - {\bf b}^S)] \left(\dfrac{2}{\pi |\lambda_{+}|}\right)^{1/2}
\: \int\limits_{0}^{\rho} \: d z \exp \left[- \dfrac{z^2}{2 |\lambda_{+}|}
 \right]
\label{Solution_FPE}
\end{eqnarray}
where we took into account the boundary condition $\pi ({\bf b}^S; {\bf u}, 0 )
= \delta ({\bf u} - {\bf b}^S)$ (vector ${\bf u} \in \partial \Omega$) and at
$\rho \rightarrow - \infty$  $\pi ({\bf b}^S) = C_{\infty}$,  i.e., well inside
the region $\Omega$  the solution is a constant as this should be for a weak
noise. In order to fix the constant $C_{\infty}$ one may multiply eq.
(\ref{In_Coordinates}) by $p_{\rm st}({\bf x})$ and after integrating over
$\Omega$ (using also the integration by parts and the Gauss theorem) derive the
surface integral
\begin{eqnarray}
 \int\limits_{\partial \Omega} \: d S  \left\lbrace - p_{\rm st}
\sum\limits_{i, j} \nu_i M_{i j}  \nabla_j H \: \pi + \varepsilon \left[
p_{\rm st} \sum\limits_{i, j} \nu_i M_{i j} \nabla_j \pi - \pi
\sum\limits_{i, j} \nu_i M_{i j} \nabla_j p_{\rm st} \right] \right\rbrace
= 0
\label{Gauss_2}
\end{eqnarray}

Given that $p_{\rm st} = Z_A^{-1} \exp [- H/\varepsilon]$, the 1-st and
3-rd terms  in eq. (\ref{Gauss_2}) cancel each other and therefore
\begin{eqnarray}
 \int\limits_{\partial \Omega} \: d S \: p_{\rm st} \: \sum\limits_{i, j} \nu_i
M_{i j} \nabla_j \pi = 0
\label{Gauss_3}
\end{eqnarray}
The gradient $\nabla_j \pi$, calculated from eq.(\ref{Solution_FPE}) at the
boundary (i.e. at $\rho = 0$), reads
\begin{eqnarray}
 \nabla_j \pi |_{\rho=0} = \nu_j \: [C_{\infty} -
\delta ({\bf u} - {\bf b}^S)] \left( \dfrac{2}{\pi
|\lambda_{+}|}\right)^{1/2}
\label{Gradient}
\end{eqnarray}
Substitution of eq.(\ref{Gradient}) into eq. (\ref{Gauss_3}) leads finally to
the result for the exit probability
\begin{eqnarray}
 \pi ({\bf b}^S) = C_{\infty} = \dfrac{\sqrt{\kappa({\bf
b}^S)}\: {\rm e}^{- H({\bf b}^S)/\varepsilon}}{
\int\limits_{\partial \Omega} \: \sqrt{\kappa ({\bf u})}{\rm e}^{- H({\bf
u})/\varepsilon}} = \dfrac{\sqrt{\kappa({\bf
b}^S)}\: {\rm e}^{- H({\bf b}^S)/\varepsilon}}{
\sum\limits_{r= 1}^{M} \: \sqrt{\kappa ({\bf b}^r)}{\rm e}^{- H({\bf
b}^r)/\varepsilon}}
\label{Exit_Probability}
\end{eqnarray}
In the last equality in eq. (\ref{Exit_Probability}) the surface integral is
replaced by a sum over all saddle (exit) points. This is possible because all
eigenvalues of the Hessian $\nabla_i \nabla_j H ({\bf u}) |_{{\bf u}={\bf b}^r}$
are positive within the boundary hypersurface $\partial \Omega$ and hence $H
({\bf u})$  has minimums at ${\bf u} = {\bf b}^r$ (where $r = 1, 2, \dots, M$).
The exit probability given by eq. (\ref{Exit_Probability}) will be compared in
Section \ref{MD_results} with our MD-simulation results.

Finally, we stress that the choice of the coordinate system given by eq.
(\ref{Coordinates}) so that only $z$-direction is physically relevant is in
complete agreement with the Kramers-Langer approach where the linear combination
(see the paragraph after eq. (\ref{FP_Harmonic})) plays the role of the relevant
coordinate. Indeed, the first relationship  from eq.(\ref{Coordinates}) can be
formally solved as
\begin{eqnarray}
 z({\bf x}) = \sum\limits_{i} \: \nu_j (x_j - b_j^S)
\end{eqnarray}
which after transformation to the principal-axis  coordinates, i.e., $\xi_i =
\sum_{j} D_{i j} (x_j - b_j^S)$ leads to $z(\{\xi_j\}) = \sum_{j} U_j \xi_j$
(where $U_j = \sum_{k} D_{j k} \nu_k$). Moreover, on the separatrix $\sum_{j}
U_j \xi_j = 0$ and one recovers the Kramers-Langer choice of a relevant
coordinate (see the paragraph after eq. (\ref{FP_Harmonic})) as linear
combination of all $\xi_j$.

\section{Breakage of an one-dimensional string of beads}
\label{Model_description}

Here we describe chain breakage by means of the Kramers approach, sketched in
Section \ref{Kramers_Langer}.

\subsection{Model}

\begin{figure}[ht]
\begin{center}
\includegraphics[scale=0.7, angle=0]{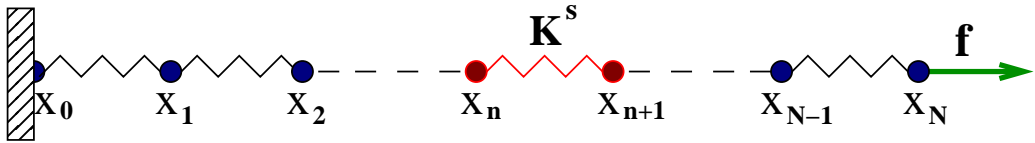}
\caption{Schematic representation of a tethered string of beads, subject to a
pulling force $f$. The corresponding coordinates are marked as $x_1, x_2, \dots
x_N$. A single ``endangered'' bond (red) with length close to the distance of
the one-bond potential maximum is located between the $n$-th and $(n+1)$-th
beads and its  spring constant $K^S$ may become negative.}
\label{Red_Bond}
\end{center}
\end{figure}

We consider a tethered one-dimensional string of $N$ beads  which experiences
a tensile force $f$ at the free end as depicted in Fig.\ref{Red_Bond}.
Successive beads are joined by bonds, governed by the Morse potential, $U_{M}(y)
= D (1 - {\rm e}^{- a y})^2$, where $D$ and $a$ are parameters, measuring the
bond strength and elasticity. The total potential energy is
\begin{eqnarray}
 V (\{x_i\}) = \sum_{i = 1}^{N} \: U_{M}(x_{i} - x_{i-1}) - f x_{N}
\label{Morse_1}
\end{eqnarray}
where we set $x_0 = 0$ (see Fig. \ref{Red_Bond}).
Upon change of variables, $y_i = x_i - x_{i-1}$, one gets
\begin{eqnarray}
 V (\{x_i\}) = \sum_{n = 1}^{N} \: \left[ U_{M}(y_n) - f y_n \right]  = \sum_{n
= 1}^{N} \: U (y_n)
\label{Morse_2}
\end{eqnarray}
so the combined one-bond potential then reads $U (y) = D (1 - {\rm e}^{- a y}
)^2 - fy$.
 
Direct analysis of the one-bond potential $U (y)$ indicates that the positions
 of the (metastable) minimum $y_{-}$ and of the maximum $y_{+}$ are given by
\begin{eqnarray}
 y_{- , +} = \dfrac{1}{a} \: \ln \Biggl[ \dfrac{2}{1 \pm \sqrt{1 -
{\tilde f}}}\Biggr]
\label{Min_Max}
\end{eqnarray}
where the dimensionless force ${\tilde f} = 2 f/a D$. The activation energy
(barrier height) is given by
\begin{eqnarray}
 E_{b} = U (y_{+}) - U (y_{-}) = D \left\lbrace \sqrt{1 - {\tilde f}} +
\dfrac{{\tilde f}}{2}  \:  \ln \Biggl[ \dfrac{1 - \sqrt{1 - {\tilde f}}}{1
+  \sqrt{1 -
{\tilde f}}}\Biggr]\right\rbrace
\label{Barrier}
\end{eqnarray}
One can easily verify that $E_b$ decreases with ${\tilde f}$. Since the Kramers'
theory implies $E_b \gg k_B T$, the force ${\tilde f}$ should not be too large.
The characteristic frequencies at the minimum and maximum of $U (y)$,
$\Omega_{1}^2 = (1/m) (d^2 U(y)/d y^2)_{y=y_{-}}$ and $\Omega_{2}^2 = - (1/m)
(d^2 U(y)/d y^2)_{y=y_{+}}$ , are therefore
\begin{eqnarray}
 \Omega_{1, 2}^2  = \dfrac{a^2 D}{m} \left[ \sqrt{1 -
{\tilde f}} \pm (1 - {\tilde f})\right].
\label{Frequency}
\end{eqnarray}

\begin{figure}[ht]
\begin{center}
\includegraphics[scale=0.2]{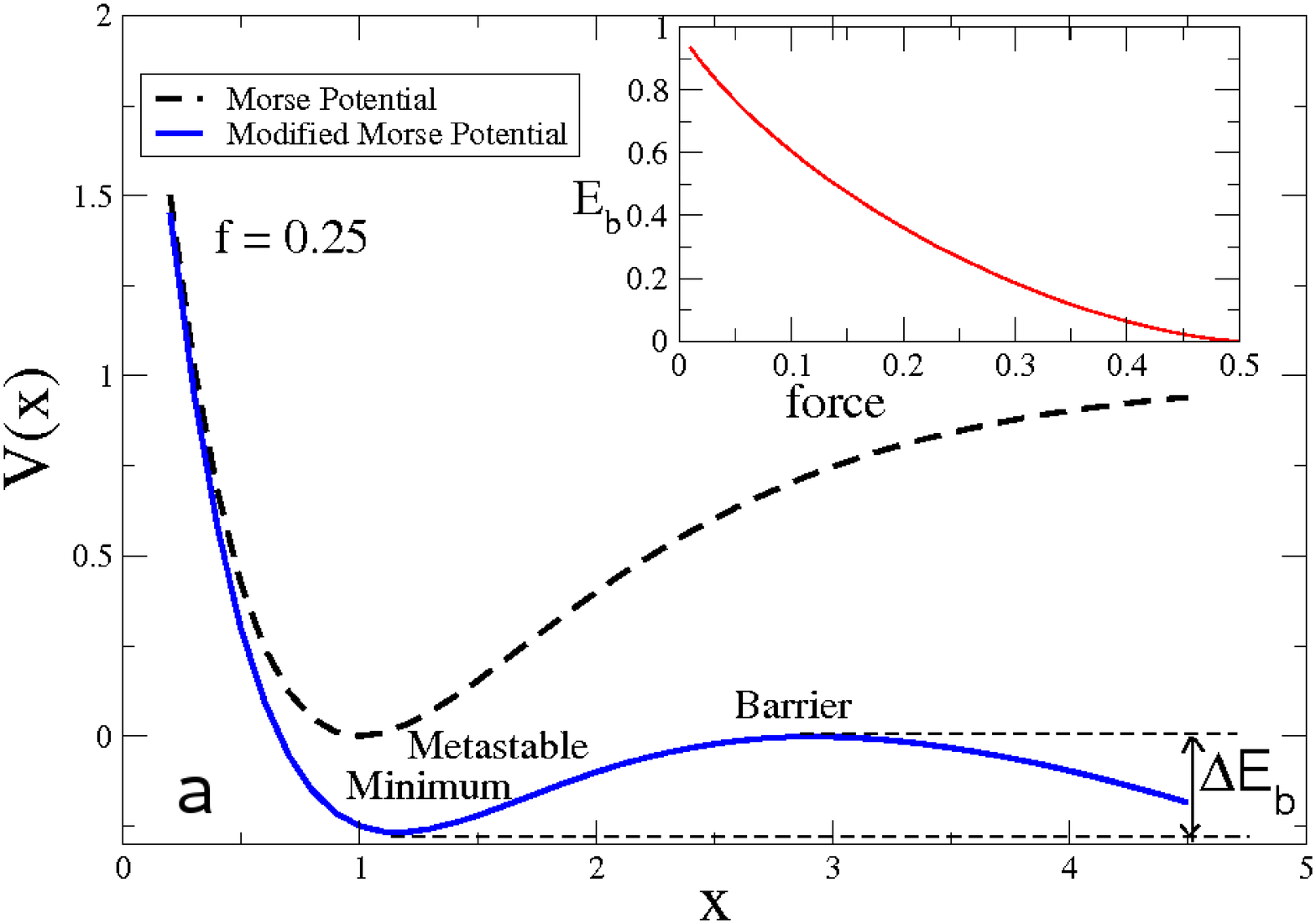}
\hspace{1cm}
\includegraphics[scale=0.2]{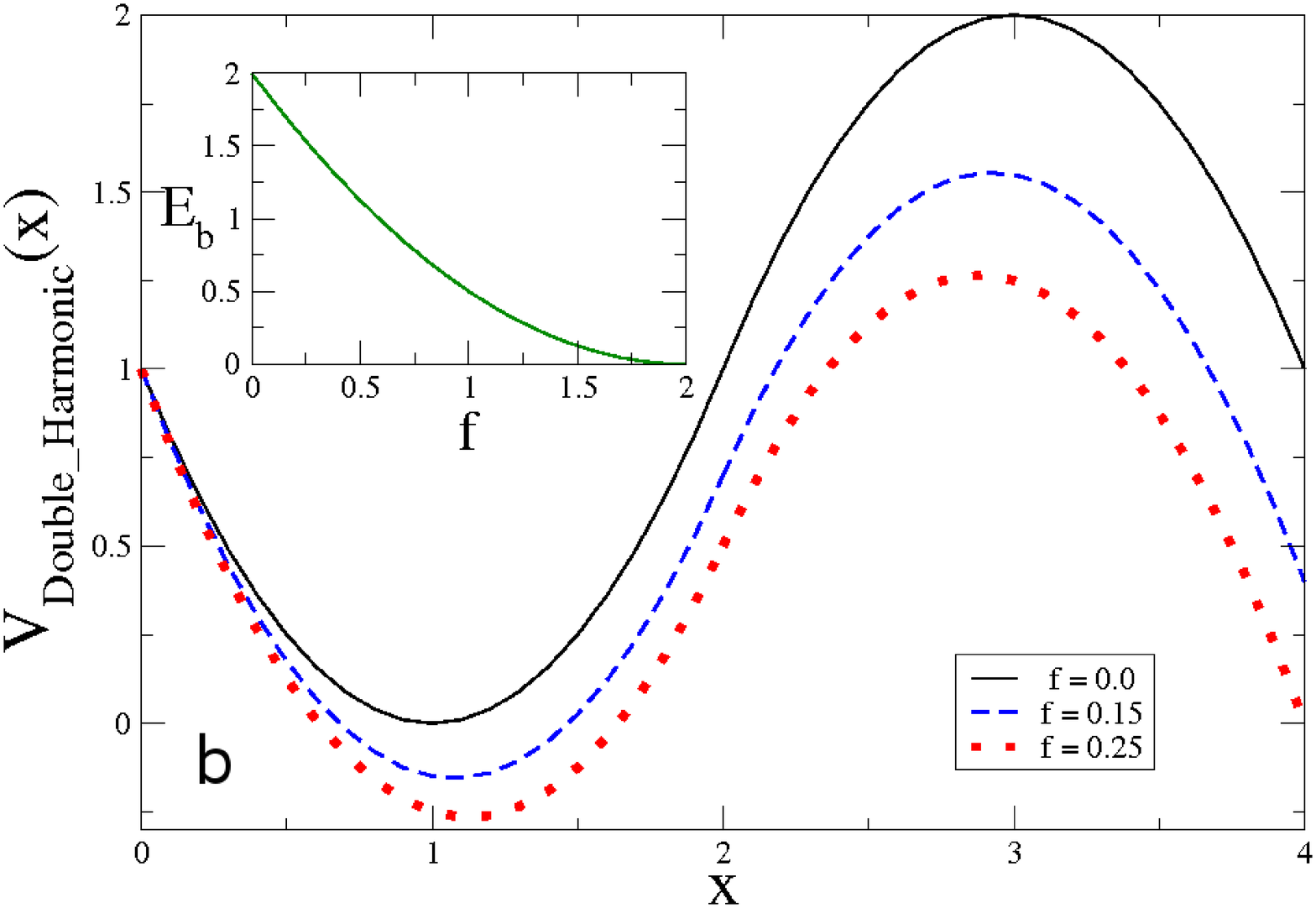}
\caption{Combined bond potential: (a) The pulling force gives rise to a
metastable minimum and a barrier $E_b$; the barrier hight $E_b$ declines with
growing force $f$ as shown in the inset. (b) The effective Double-Harmonic
potential constructed piecewise as $V(x) = (x - 1)^2 - fx$ for $x \leq 2$ and 
$V(x) = 2 - (x - 3)^2 - fx$ for $x \geq 2$ for different values of the force
$f$; the barrier hight $E_b$ declines with growing force $f$ as shown in the
inset. }
\label{Morse_Potential}
\end{center}
\end{figure}

Fig. \ref{Morse_Potential}a illustrates the Morse potential as well as the
modified one-bond Morse potential $U(y) = D(1 - {\rm e}^{-ay})^2 - f y$. In
order to test the role of anharmonicity (recall that the Kramers-Langer theory
uses only harmonic approximation) we have also tested in our MD-simulation the
Double-Harmonic potential constructed piecewise as $V(x) = (x - 1)^2 - fx$ for
$x \leq 2$ and  $V(x) = 2 - (x - 3)^2 - fx$ for $x \geq 2$ and shown in Fig.
\ref{Morse_Potential}b. The corresponding barrier heights dependences on the
pulling force are shown in inserts.

\subsection{Eigenvalues close to the metastable minimum}

One can verify that the determinants of the Hessian matrix, entering
eq.(\ref{Rate_Constant_3}), may be calculated exactly for our one-dimensional
model and even the total eigenvalue problem may be solved analytically.

The  Hessian  at the metastable minimum ${\bf x}^A$ has a $2N\times 2N$
block-matrix form
\begin{eqnarray}
 {\bf E}^A = \left.\dfrac{\partial^2 H}{\partial x_i \partial
x_j}\right|_{A} =
\left(\begin{array}{ccc}
  V_{i j}^A&   &
\begin{array}{ccc}
 0 &\cdots & 0\\
\vdots &   & \vdots\\
0 & \cdots & 0
\end{array}
\\
 \begin{array}{ccc}
 0 &\cdots & 0\\
\vdots &   & \vdots\\
0 & \cdots & 0
\end{array}&  &\begin{array}{ccc}
 1/m &\cdots & 0\\
\vdots &  \ddots & \vdots\\
0 & \cdots & 1/m
\end{array}
 \end{array}\right)
\label{Hessian_at_A}
\end{eqnarray}
where $V_{i j}^A$ is the $N\times N$-matrix of the potential energy second
derivatives. It has the following {\it tridiagonal} structure
\begin{eqnarray}
 V_{i j}^A = m \Omega_{1}^2 \left(\begin{array}{rrrrrr}
2 & -1 &  & & &  {\bf 0}\\
-1 & 2 & -1 & & &         \\
0 & -1 & 2 & -1 & &       \\
 & \ddots &\ddots  & \ddots &\ddots\\
 &  &\ 0 &-1 & 2 & -1\\
{\bf 0} &  & & &-1 &1
 \end{array}\right)
\label{V^A}
\end{eqnarray}
The correct diagonal and off-diagonal elements follow from the double
derivation of $V^A(\{x_i\}) = U(x_1) + U(x_2 - x_1) + \dots + U (x_{N-1} -
x_{N-2}) + U (x_{N} - x_{N-1})$. For example, the upper-left element $V_{1 1}^A
= \partial^2 U(x_1)/\partial x_1^2 + \partial^2 U(x_2 - x_1)/\partial x_1^2 = 2
m \Omega_1^2$ whereas the bottom-right element reads $V_{N N}^A = \partial^2
U(x_N - x_{N-1})/\partial x_N^2 = m \Omega_1^2$. The only nonzero off-diagonal
elements are $ V_{i+1 i}^A = V_{i i + 1}^A = \partial^2 U(x_{i+1} -
x_i)/\partial x_{i+1} \partial x_{i} = - m \Omega_1^2$.

The block-matrix structure ${\bf E}_{i j}^A  =
\left(\begin{smallmatrix}
       {\bf A} & {\bf 0}\\
{\bf 0} & {\bf D}
      \end{smallmatrix}\right)
$ makes it possible to calculate the determinant as \cite{Lancaster}
\begin{eqnarray}
 \det ({\bf E}_{i j}^A ) = \det {\bf A} \: \det {\bf D}
\end{eqnarray}
Thus, for the block-matrix in eq.(\ref{Hessian_at_A}) one gets
\begin{eqnarray}
\det ({\bf E}_{i j}^A ) = \left( \dfrac{1}{m}\right)^N \: \det (V_{i j}^A)\;.
\label{Det_Block}
\end{eqnarray}
The calculation of the tridiagonal matrix $V_{i j}^A$ is given in the Appendix
\ref{Appendix_2}. Using eq. (\ref{Det_Final}), one derives
\begin{eqnarray}
 \det (V_{i j}^A) = (m \Omega_1^2)^N \quad \mbox{and} \quad \det ({\rm E}_{i
j}^A) =  \Omega_1^{2N}\;.
\label{Det_V}
\end{eqnarray}

The eigenvalue problem for the Hessian eq. (\ref{V^A}) reads $V_{k n}^A
u_{n} = \lambda u_k$ , i.e., 
\begin{eqnarray}
  u_{k+1} + u_{k-1} - \left(2 - \dfrac{\lambda}{m \Omega_1^2} \right) u_k = 0
\label{E_V}
\end{eqnarray}
which should be supplemented by two  boundary conditions, namely, $u_0 = 0$
(tethered left end of the chain), and $u_N = u_{N+1}$ (free chain end right).
With the identity $ \sin (k+1) \varphi_j  +  \sin (k-1) \varphi_j - 2 \cos
\varphi_j \: \sin (k \varphi_j) = 0$, and comparing this with eq.(\ref{E_V}),
one gets for the eigenvalues  $\lambda_j = 2 m \Omega_1^2 (1 - \cos \varphi_j)$.
For the eigenfunctions one has $u_k = \sin (k \varphi_j)$ with $\varphi_j$
being the mode factor which can be fixed by the free-end boundary condition
$u_{N} = u_{N+1}$. Eventually, one obtains $\varphi_j = (2 j - 1)/(2 N + 1)
\pi$ so that the eigenvalues read
\begin{eqnarray}
 \lambda_j = 2 m \Omega_1^2 \left[1 - \cos \left(\dfrac{2 j - 1}{2 N + 1} \pi
\right)\right]\;.
\label{Eigenvalue}
\end{eqnarray}
The corresponding eigenfunctions are
\begin{eqnarray}
 u_k^{(j)} = \sin \left[ \dfrac{(2 j - 1) k}{2 N + 1} \pi \right]
\label{Eigenfunction}
\end{eqnarray}

\subsection{The determinant of the Hessian matrix at the saddle point}
\label{Three_C}

The chain breaks when at least one bond length comes close to the barrier
position $y_{+}$, eq. (\ref{Min_Max}),  of the one-bond potential $U(y)$. The
spring constant $K^S$ of this ``endangered'' bond is negative, i.e.
\begin{eqnarray}
 K^S = - m \Omega_2^2 = a^2 D \left[ - \sqrt{1 - {\tilde f}} + (1 - {\tilde
f})\right]
\label{Spring_Const}
\end{eqnarray}
Such a bond, located between the $n$-th and $(n+1)$-th beads, is illustrated  in
Fig. \ref{Red_Bond}.

As before, (cf. eq.(\ref{Det_Block})), one has
\begin{eqnarray}
 \det ({\bf E}_{i j}^S ) = \left( \dfrac{1}{m}\right)^N \: \det (V_{i j}^S)
\label{Det_S}
\end{eqnarray}
where now the Hessian of the potential energy  at the saddle point, $V_{i j}^S$,
has a more complicated  structure than $V_{i j}^A$. Indeed, from the potential
energy $V = U(x_1) + U (x_2 - x_1) + \dots + U (x_n - x_{n-1}) + U (x_{n+1} -
x_{n}) + U (x_{n+2} - x_{n+1}) + \dots + U (x_{N-1} - x_{N-2}) + U (x_{N} -
x_{N-1}) $ one recovers the following structure:  Diagonal terms: $V_{i i}^S =
2 m \Omega_1^2$ for $i \neq n, n+1, N$, $V_{n n}^S = V_{n+1, n+1}^S = m
(\Omega_1^2 - \Omega_2^2)$, and $V_{N N}^S = m \Omega_1^2$;  Off-diagonal
non-zero terms: $V_{i, i+1}^S = V_{i+1, i}^S = - m \Omega_1^2$ for $i \neq n$,
and $V_{n, n+1}^S = V_{n+1, n}^S = m \Omega_2^2$.  In result the tridiagonal
Hessian  matrix $V_{i j}^S$ reads
\begin{eqnarray}
&&\hspace{3.8cm} \downarrow{n} \hspace{0.8cm} \downarrow{n+1}\\
 V_{i j}^S &=& m \Omega_{1}^2 \left(\begin{array}{ccccccccc}
2 & -1 &  & & &  & &  & {\bf 0}\\
-1 & 2 & -1 & & &         \\
0 & -1 & 2 & -1 & &       \\
  & \ddots &\ddots  & \ddots &\ddots\\
& & 0&  -1 & 1-\alpha & \alpha\\
 & & & 0 & \alpha &1-\alpha & -1\\
& & & & \ddots & \ddots  & \ddots &\ddots\\
& &  & &  &  0 &-1 & 2 & -1\\
{\bf 0} & & & &  & & &-1 &1
 \end{array}\right)\stackrel{n}{\longleftarrow}
\label{V^S}
\end{eqnarray}
where $\alpha = \Omega_2^2/\Omega_1^2 < 1$. Let us calculate first $\det (V_{i
j}^S)$. The matrix can be considered as a block-matrix
\begin{eqnarray}
 V_{i j}^S (N, n) = m \Omega_1^2 \left(\begin{array}{cc}
                           A & B\\
C & D
                          \end{array}\right)
\label{Block_Matrix}
\end{eqnarray}
where as arguments in  $V_{i j}^S (N, n)$ we keep the total
matrix dimension, $N$, and the position of the ``endangered'' bond $n$. The
$n \times n$-block $A$, $n \times (N-n)$-block $B$, $(N-n)\times n$-block $C$
and $(N-n)\times (N-n)$-block $D$ in eq.(\ref{Block_Matrix}) are given by
\begin{eqnarray}
 A &=& \left(\begin{array}{rrrrrr}
2 & -1 &  & & &  {\bf 0}\\
-1 & 2 & -1 & & &         \\
0 & -1 & 2 & -1 & &       \\
 & \ddots &\ddots  & \ddots &\ddots\\
 &  &\ 0 &-1 & 2 & -1\\
{\bf 0} &  & & &-1 &1 - \alpha
 \end{array}\right)_{n \times n} \quad
B = \left(\begin{array}{rrr}
0 & \cdots  & 0\\
\vdots&  &\vdots\\
\alpha & \cdots   & 0
\end{array}\right)_{n \times (N-n)}\\
C &=& \left(\begin{array}{rrr}
0 & \cdots  & \alpha\\
\vdots&  &\vdots\\
0 & \cdots   & 0
\end{array}\right)_{(N-n) \times n} \qquad \qquad D =
\left(\begin{array}{rrrrrr}
1-\alpha & -1 &  & & &  {\bf 0}\\
-1 & 2 & -1 & & &         \\
0 & -1 & 2 & -1 & &       \\
 & \ddots &\ddots  & \ddots &\ddots\\
 &  &\ 0 &-1 & 2 & -1\\
{\bf 0} &  & & &-1 &1
 \end{array}\right)_{(N-n) \times (N-n)}
\end{eqnarray}
The block-matrix's determinant is given by
\begin{eqnarray}
 \det (V_{i j}^S (N, n)) = (m \Omega_1^2)^{N} \det( A) \det (\underbrace{D - C
A^{-1} B}_{F})
\label{Det_Det}
\end{eqnarray}
The $(N-n)\times (N-n)$-matrix $F = D - C A^{-1} B$ can be readily calculated to
\begin{eqnarray}
 F (N-n) = \left(\begin{array}{rrrrrr}
\chi_n & -1 &  & & &  {\bf 0}\\
-1 & 2 & -1 & & &         \\
0 & -1 & 2 & -1 & &       \\
 & \ddots &\ddots  & \ddots &\ddots\\
 &  &\ 0 &-1 & 2 & -1\\
{\bf 0} &  & & &-1 &1
 \end{array}\right)
\label{F_Matrix}
\end{eqnarray}
where $\chi_n = 1 - \alpha [1 + \alpha (A^{-1})_{n n}]$, and $(A^{-1})_{n n}$ is
the bottom-right element of the matrix $A^{-1}$. It is easy to verify that
$(A^{-1})_{n n} =  n/(1-\alpha n)$, so one gets
\begin{eqnarray}
 \chi_n = 1 - \alpha \left[1 + \dfrac{\alpha n}{1-\alpha n}\right] = 1 -
\dfrac{\alpha}{1 - \alpha n}\; .
\label{Element_chi}
\end{eqnarray}
To calculate $\det (F)$, the determinant is expanded in minors regarding 
the first row (see a similar expansion in Appendix (\ref{Appendix_2})).This
yields
\begin{eqnarray}
 \det (F (N-n)) = (\chi_n - 1)\;.
\label{Det_F}
\end{eqnarray}
On the other hand, by making use eq. (\ref{Det_Final}), we have
\begin{eqnarray}
 \det[A (n)] = (1 - \alpha n)
\label{Det_A}
\end{eqnarray}
Taking into account eqs. (\ref{Element_chi}), (\ref{Det_F}) and (\ref{Det_A})
in eq. (\ref{Det_Det}), one obtains eventually
\begin{eqnarray}
\det [V_{i j}^S(N, n)] = - \alpha  (m \Omega_1^2)^N\;.
\label{V^S_N_n}
\end{eqnarray}

The result given by eq.(\ref{V^S_N_n}) is only valid for $2 \le n \le N-2$. The
cases for $n=0, 1, N-1$ should be considered separately. The Hessians in these
cases look as follows
\begin{eqnarray}
 V_{i j}^S (N,0) &=&  m \Omega_{1}^2
\left(\begin{array}{rrrrrr}
1-\alpha & -1 &  & & &  {\bf 0}\\
-1 & 2 & -1 & & &         \\
0 & -1 & 2 & -1 & &       \\
 & \ddots &\ddots  & \ddots &\ddots\\
 &  &\ 0 &-1 & 2 & -1\\
{\bf 0} &  & & &-1 &1
 \end{array}\right) \quad V_{i j}^S (N,1) =  m \Omega_{1}^2
\left(\begin{array}{rrrrrr}
1-\alpha & \alpha &  & & &  {\bf 0}\\
\alpha & 1-\alpha & -1 & & &         \\
0 & -1  & 2 & -1 & &       \\
 & \ddots &\ddots  & \ddots &\ddots\\
 &  &\ 0 &-1 & 2 & -1\\
{\bf 0} &  & & &-1 &1
 \end{array}\right)\nonumber\\
V_{i j}^S (N,N-1) &=&  m \Omega_{1}^2
\left(\begin{array}{rrrrrrr}
2  & -1 &  & & &  &{\bf 0}\\
-1 & 2 & -1 & & &         \\
0 & -1 & 2 & -1 & &       \\
 & \ddots &\ddots  & \ddots &\ddots\\
 &  &\ 0 &-1 & 2 & -1\\
&  &  &  &-1 & 1-\alpha & \alpha\\
{\bf 0} &  & & & &\alpha &-\alpha
 \end{array}\right)
\end{eqnarray}
Direct calculation which uses the Laplace's formula for determinants (see
Appendix \ref{Appendix_2}) leads to the result
\begin{eqnarray}
 \det [V_{i j}^S (N, 0)] =  \det [V_{i j}^S (N, 1)] =  \det [V_{i j}^S (N,
N-1)] = - \alpha (m \Omega_1^2)^N
\label{Special_case}
\end{eqnarray}
Thus,  comparing eq. (\ref{Special_case}) with eq. (\ref{V^S_N_n}), one can see
that the value of $\det [V_{i j}^S (N, n)]$ does {\em not} depend on the
endangered bond index $n$. Taking into account eqs. (\ref{Det_S}),
(\ref{V^S_N_n}) and (\ref{Special_case}) leads to the final result for the
determinant of the Hessian matrix
\begin{eqnarray}
\det [{\bf E}^S (N, n)] = - \alpha \Omega_1^2\;.
\label{Det_E}
\end{eqnarray}
The determinant in eq.(\ref{Det_E}) is negative as it should.

The ratio of the fluctuating determinants $R(n) \equiv [\det{\bf
E}^A(N)/|\det{\bf E}^S (N,n)|]^{1/2}$ which  is involved in the general
expression for the rate constant, eq. (\ref{Rate_Constant_3}), is given by
\begin{eqnarray}
 R(n) = \dfrac{1}{\sqrt{\alpha}}\;.
\label{R}
\end{eqnarray}
We emphasize that the ratio of the fluctuating determinants $R(n)$
does not depends on $n$. The $n$-dependence of the total rate $k$ is present in
the $\kappa$-factor (see eq. (\ref{Rate_Constant_3}))  which will be discussed
in the next Section.

\subsection{The unstable mode}
\label{Unstable_Mode}

In an unstable equilibrium configuration when an ``endangered'' bond has a
negative spring constant there exist $N-1$  stable modes and one unstable mode.
One may find the eigenvalue $\lambda < 0$  and the eigenfunction  $u_k$
for the unstable mode.

The eigenvalue problem for the Hessian, given by eq. (\ref{V^S}), reads
$V_{k r}^S u_{r} = \lambda u_k$. In detail this yields
\begin{eqnarray}
 u_{k-1} + u_{k+1} - \left(2 - \dfrac{\lambda}{m \Omega_1^2} \right) u_k &=& 0
\quad \mbox{for} \quad k \neq n, n + 1 \label{E_P_1}\\
\nonumber\\
u_{n - 1} - \alpha u_{n + 1} - \left(1 - \alpha  - \dfrac{\lambda}{m \Omega_1^2}
\right) u_n &=& 0\quad \mbox{for} \quad k = n \label{E_P_2}\\
\nonumber\\
 - \alpha u_{n} +    u_{n + 2} -  \left(1 - \alpha  - \dfrac{\lambda}{m
\Omega_1^2}
\right) u_{n + 1} &=& 0 \quad \mbox{for} \quad k = n + 1
\label{E_P_3}
\end{eqnarray}
which should be supplemented by the boundary conditions: $u_0 = 0$ (tethered
end of the polymer) and $u_{N} = u_{N+1}$ (free end of the chain). We recall
that the index $n$ denotes an ``endangered'' bond location between the $n$-th
 and $(n+1)$-th beads.

Let us first find the solution for $k < n$. To this end we use the identity $
\sinh [(k+1)\varphi] + \sinh [(k-1)\varphi]  - 2 \cosh (\varphi) \sinh ( k
\varphi) = 0 \label{Identity}$. Comparison of this identity  with eq.
(\ref{E_P_1}) suggests that the eigenfunction which characterizes the distance
from the unstable equilibrium position is given by
\begin{eqnarray}
u_k = - \sinh (k \varphi)
\label{u_k}
\end{eqnarray}
with an eigenvalue
\begin{eqnarray}
 \lambda = 2 m \Omega_1^2 (1 - \cosh \varphi)
\label{Lambda_Unstable}
\end{eqnarray}
where $\varphi$ is the mode factor which will be fixed below. The solution
eq.(\ref{u_k}) also meets the boundary condition $u_0 = 0$.

In order to find the solution at $k > n + 1$ we consider two identities
\begin{eqnarray}
 \sinh[(N + 1 - k-1)\varphi] + \sinh[(N + 1 - k + 1)\varphi] - 2 \cosh
(\varphi) \sinh[(N + 1 - k)\varphi] &=& 0\\
\cosh[(N + 1 - k-1)\varphi] + \cosh[(N + 1 - k + 1)\varphi] - 2 \cosh
(\varphi) \cosh[(N + 1 - k)\varphi] &=& 0
\label{Identity_2}
\end{eqnarray}
Again, comparison with eq. (\ref{E_P_1}) suggests that there
are two linearly  independent, i.e., fundamental solutions ( see, e.g., \cite{
Elaydi}), $u_{k}^{(1)} = \sinh [(N+1 - k) \varphi]$ and $u_{k}^{(2)} = \cosh
[(N+1 - k) \varphi]$. Thus the general solution reads: $u_k = A \sinh
[(N+1 - k) \varphi] + B \cosh [(N+1 - k) \varphi]$ where $A$ and $B$ are some
constants. The boundary condition $u_{N+1} = u_N$ helps to express $B$ in
terms of $A$ which gives
\begin{eqnarray}
 u_k = A \left\lbrace \sinh [(N+1 - k) \varphi] - \coth
\left(\dfrac{\varphi}{2}\right) \cosh [(N+1 - k) \varphi]\right\rbrace
\label{Function_k>n}
\end{eqnarray}

We extend the solutions, given by eq. (\ref{u_k}) and eq. (\ref{Function_k>n}),
up to $k = n$ and $k = n + 1$, respectively. Consequently,
\begin{eqnarray}
 u_k = \begin{cases} - \sinh (k \varphi) , &\mbox{at}  \quad 1\leq k \leq n \\
A \left\lbrace \sinh [(N+1 - k) \varphi] - \coth
\left(\dfrac{\varphi}{2}\right) \cosh [(N+1 - k) \varphi]\right\rbrace ,
&\mbox{at} \quad n + 1\leq k \leq  N
\end{cases}
\label{Function_General}
\end{eqnarray}
The mode factor $\varphi$ and the amplitude $A$ are determined  by the
conditions eq. (\ref{E_P_2}) and eq. (\ref{E_P_3}). The substitution of eq.
(\ref{Function_General}) in eqs. (\ref{E_P_2}) and (\ref{E_P_3}) yields
\begin{eqnarray}
 \sinh [(n - 1) \varphi] &+& \alpha A \left\lbrace \sinh [(N - n) \varphi] -
\coth
\left(\dfrac{\varphi}{2}\right) \cosh [(N-n) \varphi]\right\rbrace + [1 +
\alpha - 2 \cosh (\varphi)] \sinh (n \varphi) = 0 \nonumber\\
\alpha \sinh (n \varphi) &+& A \left\lbrace \sinh [(N-n-1) \varphi] -
\coth\left(\dfrac{\varphi}{2}\right) \cosh [(N - n - 1)
\varphi]\right\rbrace\nonumber\\
&+& A [1 +\alpha - 2 \cosh (\varphi)]
\left\lbrace \sinh [(N-n) \varphi] -
\coth\left(\dfrac{\varphi}{2}\right) \cosh [(N-n) \varphi]\right\rbrace = 0
\label{Mode_Factor_1}
\end{eqnarray}

At $n = 0$ the first equation in eq.(\ref{Mode_Factor_1}) becomes redundant
and the second one can be written as
\begin{eqnarray}
\left\lbrace \sinh [(N-1)\varphi] - \coth \left(\dfrac{\varphi}{2}\right)
\cosh[(N-1)\varphi]\right\rbrace + [1+\alpha - 2\cosh \varphi] \left\lbrace
\sinh (N\varphi) - \coth \left(\dfrac{\varphi}{2}\right)\cosh
(N\varphi)\right\rbrace  = 0
\label{Mode_Factor_2}
\end{eqnarray}
For the values $1\leq n \leq N-1$ the amplitude $A$ may be excluded from eq.
(\ref{Mode_Factor_1}) which leads to the equation
\begin{eqnarray}
\alpha^2 \sinh (n\varphi) &=& \left\lbrace \sinh [(n-1)\varphi] + [1+\alpha -
2\cosh \varphi] \sinh (n \varphi)\right\rbrace \nonumber\\
&\times& \left\lbrace \dfrac{\sinh [(N-n-1)\varphi]-
\coth(\varphi/2) \cosh [(N-n-1)\varphi]}{\sinh [(N-n)\varphi]-\coth(\varphi/2)
\cosh [(N-n)\varphi]} + [1+\alpha -
2\cosh \varphi] \right\rbrace
\label{Mode_Factor_3}
\end{eqnarray}
The substitution of $n=0$ in eq. (\ref{Mode_Factor_3}) gives back eq.
(\ref{Mode_Factor_2}) as required by consistency.

From the solution of the transcendental eq.(\ref{Mode_Factor_2}) (for $n=0$) or
eq.(\ref{Mode_Factor_3}) for $1\leq n \leq N-1$ one obtains the mode factor
$\varphi$ as a function of $n$ and $N$, i.e. $\varphi (N, n)$. Knowing
$\varphi$, one may calculate the negative eigenvalue given by
eq.(\ref{Lambda_Unstable}). Making use of this in the equation for the factor
$\kappa$ (see eq. (\ref{Kappa})) one obtains
\begin{eqnarray}
 \kappa (N, n) = \dfrac{\gamma}{2} \left[ 1 - \sqrt{1 + \dfrac{8
\Omega_1^2}{\gamma^2} [\cosh \varphi(N, n) - 1]} \right] < 0
\label{Kappa_Eqation}
\end{eqnarray}

Finally, the first equation in (\ref{Mode_Factor_1}) makes it possible to
calculate the amplitude $A$ as a function of $N$ and $n$, i.e.
\begin{eqnarray}
A(N, n) = - \dfrac{\sinh [(n-1) \varphi] + [1 + \alpha - 2 \cosh
(\varphi) ] \sinh (n \varphi)}{\alpha \{\sinh [(N-n)\varphi]-\coth(\varphi /2)
\cosh [(N-n)\varphi]\}}
\label{Amplitude}
\end{eqnarray}

As mentioned in Sec. \ref{Three_C}, the chain breaks when at
least one bond approaches the position of the potential maximum, i.e. becomes
``endangered''. On the other hand, within the Kramers approach this is a
relatively rare event, so that a simultaneous occurrence of a second, third,
etc. ``endangered'' bonds may be neglected.  Therefore, by
calculating the  rate constant for the total chain one should average over all
possible locations
of an ''endangered`` bond. Consequently, the rate constant of the total
chain (see
eq. (\ref{Rate_Constant_3}) is given as
\begin{eqnarray}
 k = \dfrac{1}{2 \pi} \: {\rm e}^{-\beta E_b} \: R  \: \dfrac{1}{N}
\sum_{n=0}^{N-1} \:
|\kappa(N,
n)|
\label{Total_Rate}
\end{eqnarray}
where $R$ and $\kappa (N, n)$ are given by eqs. (\ref{R}) and
(\ref{Kappa_Eqation}), respectively.
By means of a MD-simulation, one may calculate the average first passage time
$\tau$ for crossing the barrier which is related  in turn as $k = (2
\tau)^{-1}$ to the rate constant (see Sec. \ref{FPT_Approach}  and ref.
\cite{Talkner}).

\section{Simulation Results}
\label{MD_results}

We present here the results our extensive MD simulations in order to verify the
theoretical predictions and explore the limitations of the analytical treatment.
Energy is given in units of $D$ and length is measured in units of $1/a$, the
parameters of the Morse potential  $U (y) = D (1 - {\rm e}^{- a y} )^2$. Mass is
measured in terms of $m$, the mass of the bead. Time is measured in units of $
a^{-1}\sqrt{m/D}$; temperature is measured in units of
$D/k_B$ where the Boltzmann constant, $k_B$ has been set equal to $1.0$.
Unless otherwise mentioned, the ratio of the barrier height to temperature $
E_{b}/T$ has been set to $5$, the value of the externally applied pulling force
has been set equal to $0.25$ and the friction coefficient of the Langevin
thermostat, used for equilibration, has been set at $0.25$. The integration step
is $0.002$.
 \begin{figure}[ht]
\includegraphics[height = 1.0 cm, width = 6.0 cm]{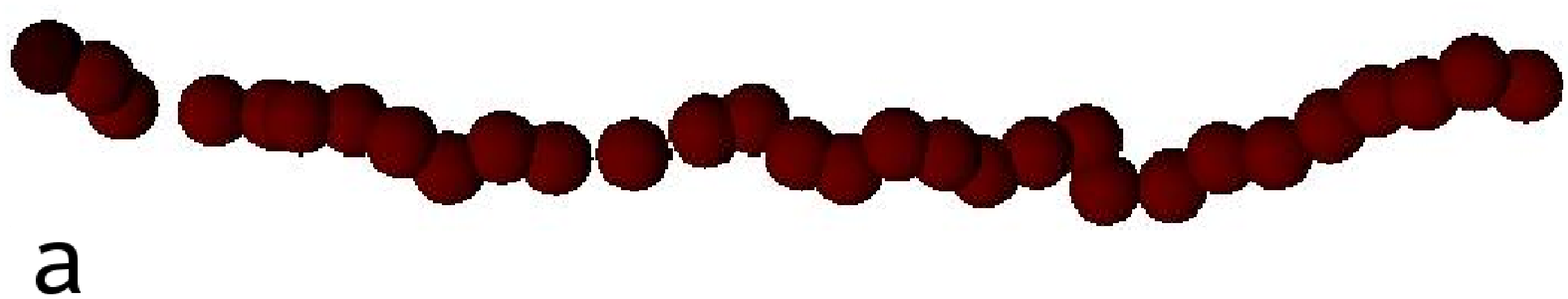}
\hspace{2.0 cm}
\includegraphics[height = 1.0 cm, width = 6.0 cm]{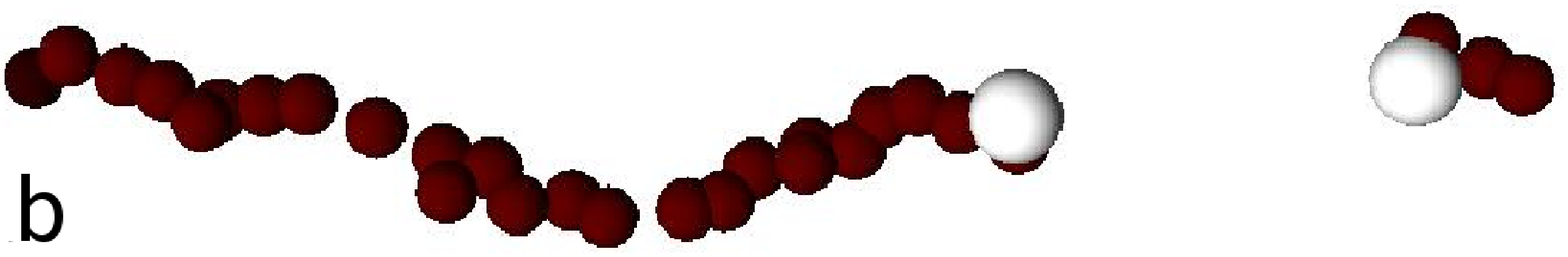}
\caption{Snapshots of a chain with $30$ beads fixed at the left end while
the right end is pulled by a constant force: (a) an equilibrated initial
conformation, (b) a broken chain with the beads at the scission site shown
in white.}
\label{Snapshot}
\end{figure}

 We start the simulation with each bead separated by a distance equal to the
equilibrium separation of the effective bond potential $U(y_n)$ (eq. (\ref
{Morse_2})); we then let the chain equilibrate with its environment using a
Langevin thermostat. The number of integration steps for equilibration of a
chain with $10$ atoms in $1$-D is $20000$ and the number of equilibrating
integration steps is increased linearly as the chain length is increased.
Because of the presence of the external pulling force, the stretched chain
attains equilibrium only locally and not globally, i.e., it never turns into a
coil. This justifies the linear increase in the number of equilibration steps
with increase in chain length as opposed to ``quadratic'' as in the Rouse
model. Once equilibration is achieved, time is set to zero and one measures the
elapsed time (in MD time units) before any of the bond lengths extends beyond
the distance separating the metastable minimum from the maximum, i.e., until one
of the beads crosses the barrier (see Fig. \ref{Snapshot}). We repeat the above
procedure for a large number of events ($5\times 10^4 \div 2\times 10^7$) so as
to sample the stochastic nature of rupture and calculate properties like the
mean rate of rupture, the distribution of breaking bonds regarding their
position in the chain, the (First Passage Time) FPT distribution, etc., for
chains of different length in both $3d$ and $1d$. As one of our principle
objectives, we also investigate the issue of chain recombination (self-healing)
which is frequently observed after a scission event occurs. We demonstrate that
the mere barrier crossing is not a reliable criterion for chain breakage since
the majority of broken bonds are observed to recombine. Therefore we develop an
unambiguous criterion for true rupture as illustrated in a later subsection.

\subsection{Chain Scission - Simulation Results}

We compare here the simulation results with our theoretical prediction for the
rupture probability of the $n$-th bond in a chain with $N$ bonds. The
probability for bond scission (or, exit probability, in the more general
context of Section \ref{Kramers_Langer}), is given by eq. (\ref
{Exit_Probability}).

 \begin{figure}[ht]
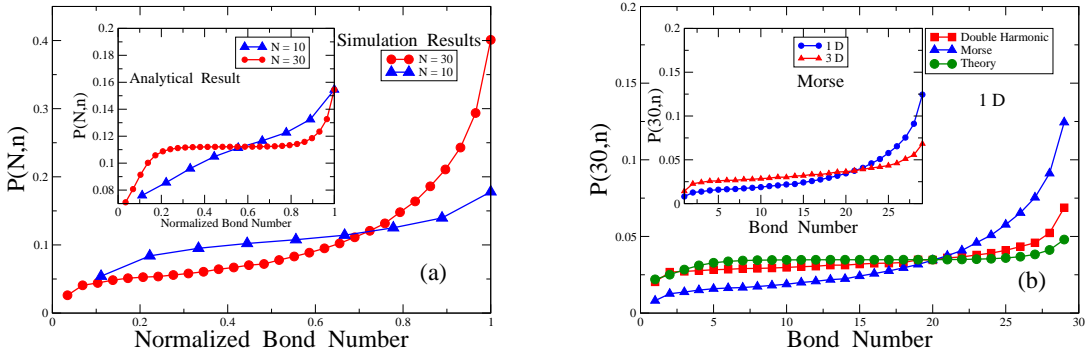

\vspace{0.50cm}
\includegraphics[scale=0.27]{fig5a.eps}
\hspace{1.0 cm}
\includegraphics[scale=0.27]{fig5b.eps}
\caption{Normalized rupture probability vs consecutive bond number for chains
with length $N$, subject to tensile force $f = 0.25$, and friction
$\gamma = 0.25$: (a) $1D$ results for $E_b/k_BT = 5$ and $N=10,\;30$. The
consecutive number of the bonds is normalized as $n/N$ for convenience. Insert
shows the theoretical prediction.
(b) Theoretical prediction for $N=30$ along with simulation results for the
Morse potential and the Double Harmonic potential. Inset shows the
same for a chain with Morse interactions in both 1D and 3D.}
\label{Rupture_Probability}
\end{figure}

In Fig. \ref{Rupture_Probability} the normalized rupture probability  for chains
(with $N = 10$ and $N = 30$) is shown with respect to the consecutive number of
the individual bonds. The theoretical prediction, which follows from the
numerical solution of eqs. (\ref{Mode_Factor_2}), (\ref{Mode_Factor_3}) and
(\ref{Kappa_Eqation}), is given in the inset. Both the theory- and MD-results
indicate that the pulled end of the chain and the bonds in its vicinity break
more frequently due to more freedom than those around the fixed end.
Generally, the probability of rupture decreases steadily from the pulled end to
the fixed end. For the longer chain, the end effects are not felt by the middle
part of the chain and the probability of rupture $P(N,n)$ is nearly uniform
forming a plateau-like region all over the length of the chain except at the
ends. This feature is more pronounced in the theoretical rather than in the MD
results. Such a comparison between the theory and MD-simulation findings has not
been done before and could be viewed as a test for the merits and
shortcomings of Kramers approach. Moreover, we believe that
this reflects the incorporation of the collective unstable mode given in Sec.
\ref{Unstable_Mode}.

 One may assume that the detected discrepancy between theory and MD-simulation
can be ascribed to the use of harmonic approximation around the
metastable minimum and unstable maximum of the effective bond potential. To
show this, we replaced the effective potential by a Double Harmonic potential -
a parabola, and an inverted parabola, which approximates the effective potential
in shape. The comparison of the three results - theoretical prediction,
simulation with Morse potential, and simulation with the Double Harmonic
potential is shown in Fig. \ref{Rupture_Probability}b. Evidently, the rupture 
probability distribution for the Double Harmonic potential exposes also a
plateau-like region and matches the theoretical prediction quite
closely. So we infer that the absence of the plateau in case of the Morse
potential can be ascribed to its anharmonicity. In the inset of
Fig. \ref{Rupture_Probability}b, we show the simulation results with the Morse
potential for a $30$-atom chain in $1D$ and $3D$. Interestingly, one may
see that the MD-results in $3D$ indicate a flatter distribution $P(N,n))$
than in $1D$, coming thus closer to theoretical predictions. As far as in $3D$
there exist much more configurations with endangered bonds due to transversal
displacements of the beads, bond length fluctuations are suppressed as compared
to $2D$ and the bond anharmonicity is less pronounced.

\begin{figure}[htb]
\includegraphics[scale=0.27]{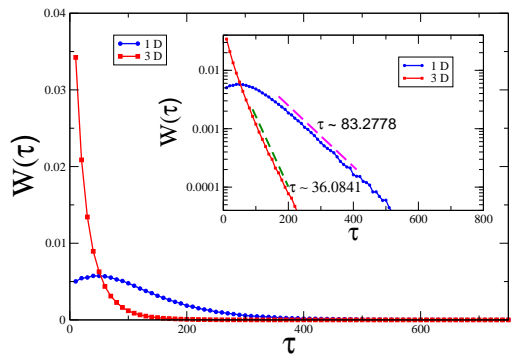}
\caption{First passage time distributions $W(\tau)$ against elapsed time
$\tau$ in $1D$ and $3D$ for a chain with $N = 30$, $f = 0.25$ and $\gamma =
0.25$. The inset shows the same in semilog coordinates. }
\label{Life}
\end{figure}

Theory predicts that the first passage time distribution goes asymptotically  as
$W \sim \exp(-t/\tau)$ ( see eq.(\ref{PDF})). In Fig. \ref{Life}, we plot the
FPT distribution for a chain with $30$ beads. One should note here the
considerable difference of $W(\tau)$ between $1D$ and $3D$. The long time tail
of the distribution is indeed seen to decay exponentially. We estimate the mean
FPT (for the whole chain and not of an individual bond) as $\tau_{1D} = 83.2$
and $\tau_{3D} = 36.0$. The theoretical estimate for a $30$-bead chain in
$1D$ is $\tau_{1D} \simeq  866.5$ which is an order-of-magnitude larger than
that estimated from the simulations. This finding is in agreement with the
results of Sain et al. - cf. Table I in \cite{Sain}.

While assessing these results one should bear in mind that, as mentioned above,
an event of bond scission can be defined as ''barrier crossing'' whereby an
endangered bond stretches beyond the position of the maximum of our potential
$U(y)$. As has been discussed in the previous section, this position of the
barrier depends on the externally applied force once the parameters of the Morse
potential are fixed. For an applied force of magnitude $0.25$, the barrier
position corresponds to a critical bead-bead separation (bond length) of
$2.92$. Since frequently such scission event is immediately followed by
recombination (that is, by self-healing of the bond), in what follows we have
also considered a more stringent criterion for irreversible rupture. As a
possible choice for the critical bead-bead separation one may take the larger
value of $5$ which renders self-healing events virtually improbable (see below).

\begin{figure}[ht]
\vspace{1.0cm}
\centering
\includegraphics[scale=0.27]{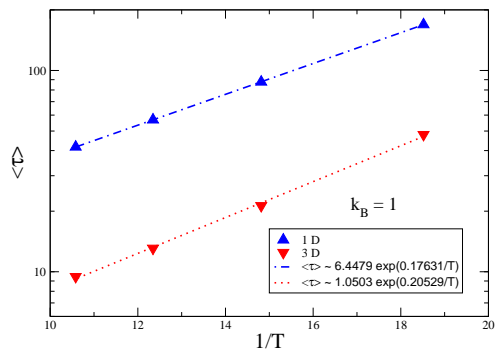}
\vspace{0.50cm}
\caption{Mean first passage time $\langle\tau\rangle$ vs inverse temperature for
a chain with $N=30$, $f=0.25$ and $\gamma = 0.25$.}
\label{Arrhenius}
\end{figure}

It is of interest to examine the dependence of mean FPT $\langle\tau\rangle$ on
temperature $T$ for a given applied force $f$. In Fig.\ref{Arrhenius} we show
the change in FPT  for irreversible rupture (and not barrier crossing) on
inverse temperature for $f = 0.25$ and $\gamma = 0.25$. As expected from
eq.(\ref{Total_Rate}), the Arrhenian nature of this relationship is clearly
manifested in a semilog plot. From the exponentially fitted values, the
effective barrier height is estimated to be $0.176$ in $1$D and $0.205$ in
$3$D. The theoretical estimate of the barrier for a force of magnitude $0.25$ is
$\Delta E_b \sim 0.27$. The somewhat lower value of the effective barrier has
also been observed earlier in simulations with the Lennard-Jones potential for
the fixed strain ensemble \cite{Oliveira_1}.

\begin{figure}[ht]
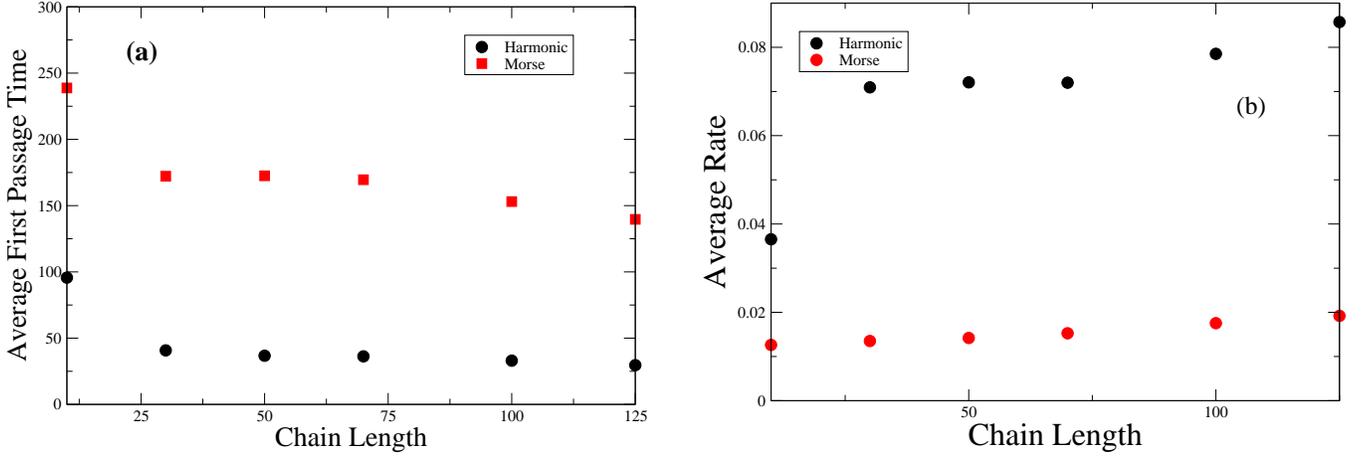

\vspace{0.9cm}
\centering
\includegraphics[scale=0.35]{fig8a.eps}
\hspace{0.5cm}
\includegraphics[scale=0.35]{fig8b.eps}
\caption{Mean first passage time $\langle\tau\rangle$ (a) and average scission
rate (b) vs chain length $N$ for Morse and double-harmonic chains with $f=0.25$
and $\gamma = 0.25$.}
\label{N_dependence}
\end{figure}

Eventually, in Fig. \ref{N_dependence} we show the $N$-dependence of the mean
FPT $\langle \tau \rangle$ and the average scission rate for both the Morse
chain and the double-harmonic chain. We recall that we considered the unstable
collective mode in Section \ref{Model_description}D as the one responsible for 
the rupture event. Therefore, one might expect that the $N$-dependence of the
scission rate would clearly reveal the degree of collectivity in bond
breaking. Indeed, a linear increase in the scission rate with growing
chain length would indicate the {\em independent} character of scission events,
that is, the breakage of a single bond occurs irrespective of the
neighboring bonds. In contrast, if the  nature of bond scission is a strongly
{\em collective} process, this $N$-dependence should be negligible. Evidently,
Fig. \ref{N_dependence} is a clear manifestation of the latter, contrary to an 
earlier assumption \cite{Oliveira_4,Sain,Sokolov_1,Sokolov_2} that the
total probability for scission of polymer with $N$ bonds is $N$ times that of a
 single bond.  One sees a negligible decline in the mean FPT  in Fig.
\ref{N_dependence}a which appears as a very weak rise in the total rate  in
Fig. \ref{N_dependence}b, much weaker (with slope $\approx 0.1$) than the
presumed linear growth with $N$.
A visible finite size effect is detected only for $N=10$, indicating a 
strong influence of the boundaries (i.e., of the chain ends). This effect is
easily understood from the inset to Fig. \ref{Rupture_Probability}a where the
absence of a plateau suggests that the bonds in the short chain are not
equivalent. One can also verify from Fig. \ref{N_dependence} that the
double-harmonic chain breaks faster than the Morse one,

\subsection{The breaking of a bond}

We now examine the expansion rate of the breaking bond. Theory predicts that the
length of the endangered bond extends with time as $\exp(-\kappa t)$ (see eq.
(\ref{Solution})), where $\kappa$, the {\em transmission} factor given by eq.
(\ref{Kappa_Eqation}), is negative.
\begin{figure}[ht]
\includegraphics[scale=0.35]{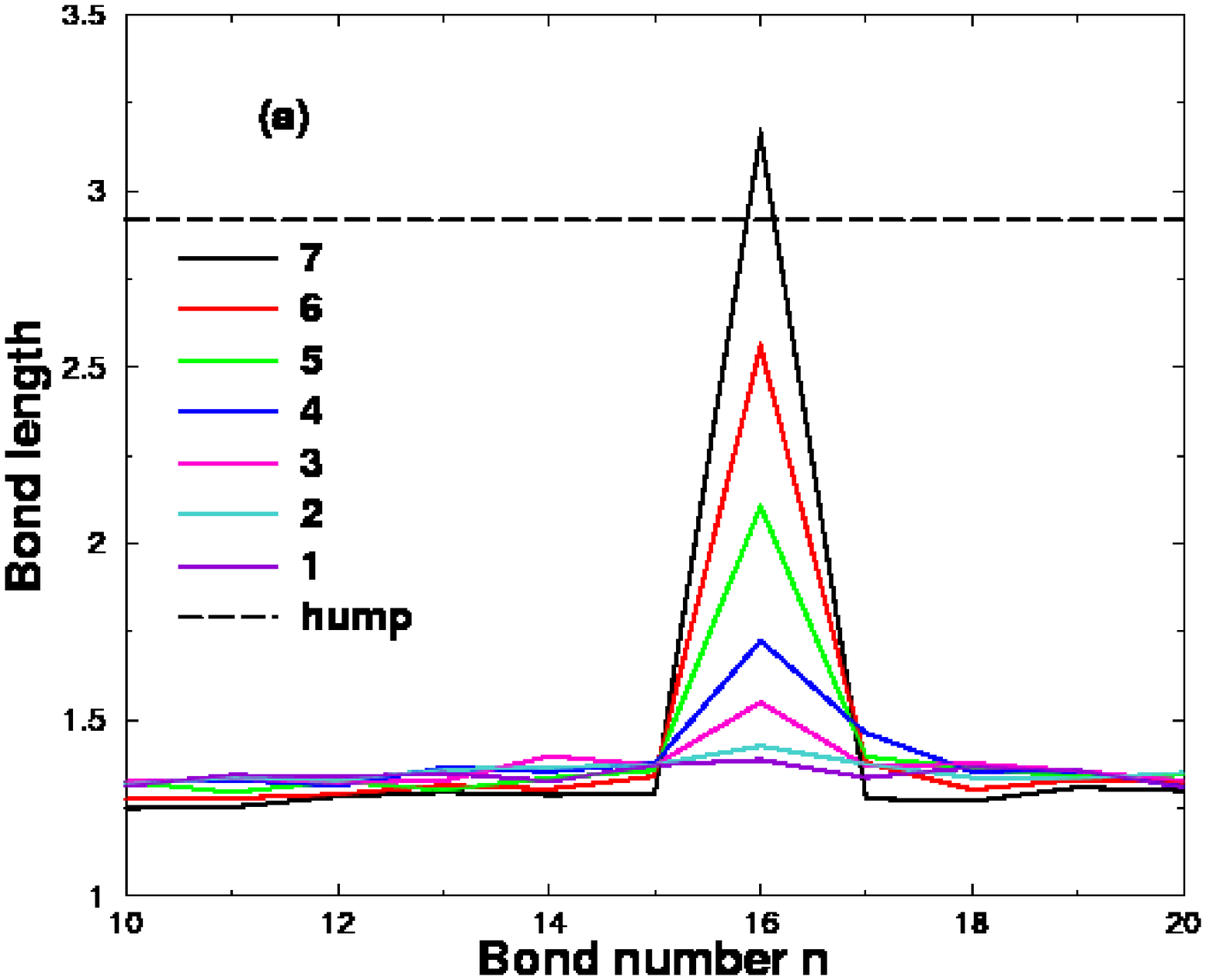}
\hspace{0.5cm}
\includegraphics[scale=0.25]{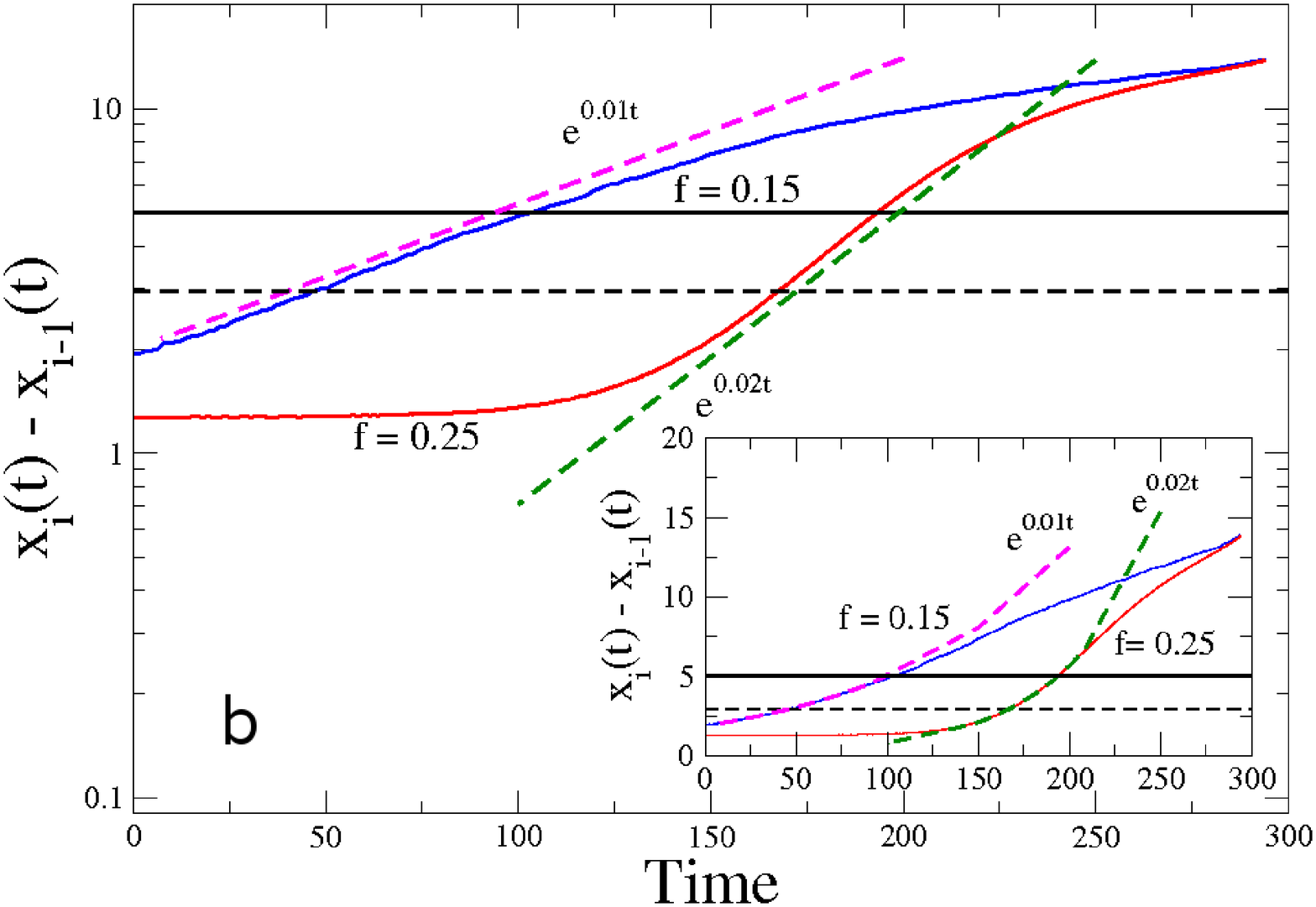}
\caption{Stretching of a breaking ($15$-th) bond with time in a chain with
$30$-beads  for $f = 0.15$ and $f = 0.25$: (a) Evolution of the lengths of a
breaking bond and of its neighbors. The dashed line denotes an extension equal
to the barrier position. (b) Variation of the length of a breaking bond with
time in semi-logarithmic coordinates. Dashed lines indicate exponential increase
in agreement with theoretical predictions. The inset shows the expansion in
normal coordinates for visual aid. The length growth has been monitored up to an
expansion of $15$, well beyond the location of the barrier. Data has been
averaged over many events. The dashed black line shows the position of the barrier while the solid line marks an expansion of 5.0, later set as the criterion for irreversible rupture for an applied force $f = 0.25$.}
\label{Expansion}
\end{figure}

\begin{figure}[ht]
\centering
\includegraphics[scale=0.27]{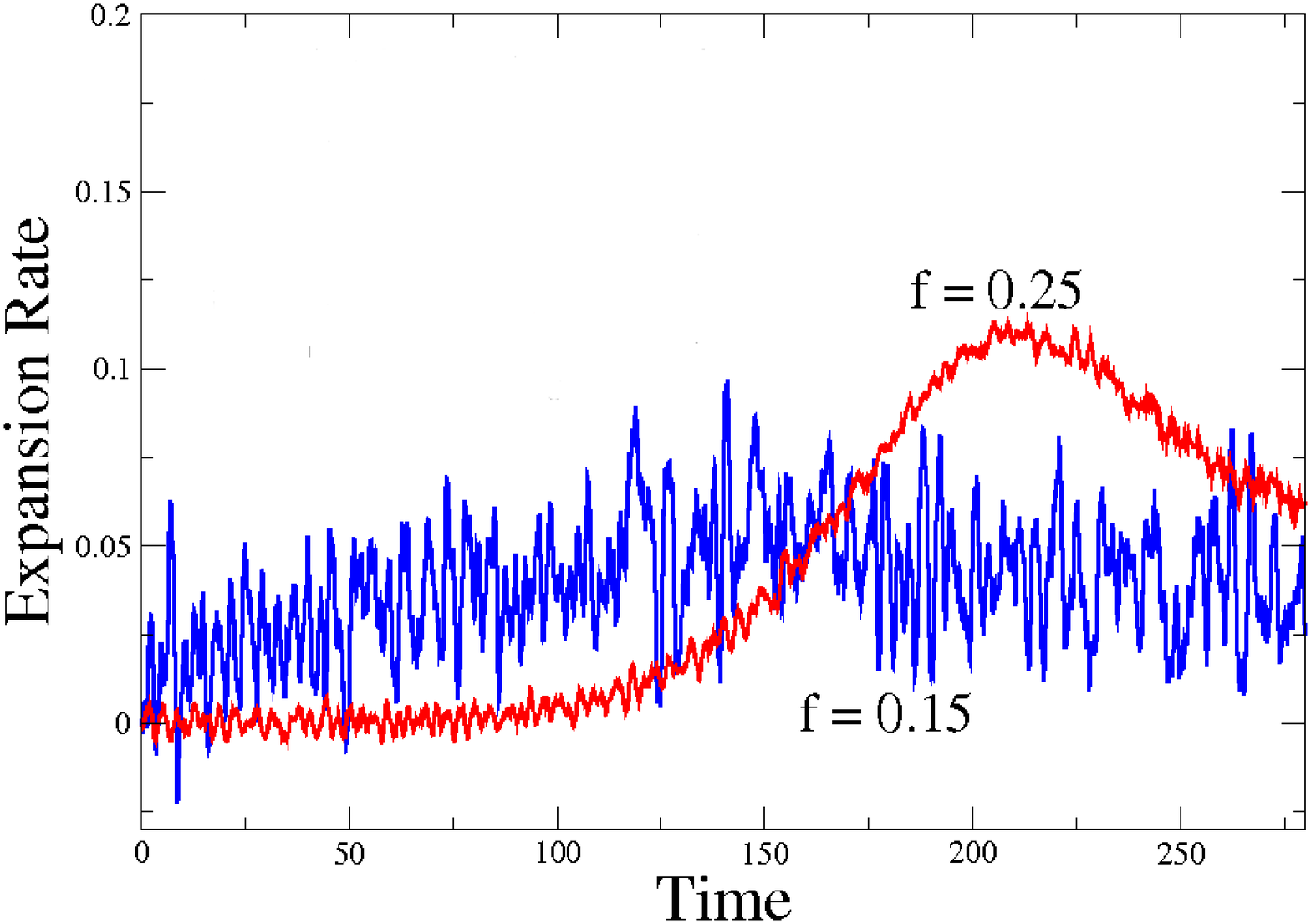}
\caption{Mean expansion rate of the breaking $15$-th bond.}
\label{Rate}
\end{figure}

Exponential fits of the growth curves around the position of the barriers ($\sim
2.92$ for $f = 0.25$ and $\sim 3.05$ for $f = 0.15$) give $|\kappa| \sim 0.01$
for $f = 0.15$ and $|\kappa| \sim 0.02$ for $f = 0.25$ is given in
Fig.\ref{Expansion}. Analytical calculations yield $0.11$ and $0.22$ - Fig.
\ref{Expansion} - respectively, an order of magnitude higher than that
estimated from the simulations and consistent with the result obtained earlier
from the lifetime distribution.

Eventually, we show the rate of expansion of the monitored bond for two
different forces, note that the exponential character of the curves is best
pronounced when the growing bond length is in the vicinity of the barrier
(hump) position which is indicated in Fig. \ref{Expansion} by horizontal lines,
in agreement with the theoretical predictions, eq.(\ref{Solution}). The speed of
expansion itself is displayed in Fig. \ref{Rate} (statistic fluctuations are
strongly pronounced for the weaker tension $f=0.15$) which demonstrates that
also the speed of bond extension attains a maximum at the moment it goes over
the barrier position.

\section{Self-Healing (Recombination) of Broken Bonds}
\label{Self_Healing}

The process of bond breakage is not fully described by the escape from the
metastable well and the barrier crossing. Once the barrier has been surmounted,
the ''endangered bond'' could further increase its length up to a critical
value which still leaves the possibility for eventual return and healing of the
chain. In contrast, beyond this critical value the healing probability gets
negligibly small and the chain breaks irreversibly. This healing process has
been pointed out earlier \cite{Sebastian} and observed in a MD-simulation
\cite{Bolton}. Below we suggest a theoretical description of the process of
recombination and derive analytical expressions for the healing time and
distance distribution function. One should bear in mind that it is the thermal
fluctuations that initiate recombination events while the downhill ramp
potential of the applied tensile force always acts in the opposite
''destructive'' direction. This potential may still be approximated by an
inverted parabola before the ''endangered bond'' length approaches a critical
size whereby the two pieces of the chain would move apart irreversibly.

\subsection{Theory}

As far as the healing process is largely localized on the ''endangered bond'',
it may be treated as an effectively one-dimensional problem. The corresponding
Fokker-Planck equation for the probability distribution function of positions
(in our case - ``endangered'' bond lengths) and velocities (bond length
velocities) is known as Kramers equation\cite{Kramers} and has been used to
describe reaction kinetics. The complete solution of the Kramers equation for
the harmonic as well as inverted harmonic (or inverted parabolic) potential
was Fig. \ref{N_dependence}
given by Risken \cite{Risken}.

Consider the Kramers equation for the inverted harmonic potential $U(x) = -
\Omega_2^2 x^2/2$, assumed to represent the top of the barrier. The
coordinate-velocity transition probability  $P (x, v, t| x', v', 0)$ is governed
by the following equation (see Sec. 10 in \cite{Risken}):
\begin{eqnarray}
 \dfrac{\partial}{\partial t} \: P = -
\dfrac{\partial}{\partial x} [v P] +
\dfrac{\partial}{\partial v} \: [(U'(x) + \gamma v )  P] +
\gamma \: v_{\rm th}^2 \: \dfrac{\partial^2}{\partial v^2} \: P
\label{Kramers_Equation}
\end{eqnarray}
where the thermal velocity $v_{\rm th} = \sqrt{T/m}$  and the initial conditions
are fixed as $P (x, v, t=0| x', v', 0)= \delta (x-x') \: \delta (v - v')$.
The general solution of eq. (\ref{Kramers_Equation}) is given by \cite{Risken}
and is relegated to Appendix \ref{K_Equation_Solution}.

\subsubsection{Healing time distribution function} 

One may treat the healing time distribution function by noting that the process
starts at the top of the inverted harmonic potential $U(x) = - \Omega_2^2
\:x^2/2$, where $x' = 0$ and $v' = |\kappa|$ . The coordinate of the turning
point (healing distance) is counted with respect to $x = 0$ whereas the velocity
at that point might be arbitrary within the interval $- \infty \leq v \leq 0$
(i.e., the return velocity is pointing to the left). Thus, the healing time PDF
can be obtained from the transition probability eq. (\ref{General_Solution}) as
\begin{eqnarray}
P_{\rm heal} (t) = \int\limits_{- \infty}^{0} \: d v \:P (0, v, t| 0, |\kappa|,
0)
\label{Integration}
\end{eqnarray}
After explicit integration in eq. (\ref{Integration}) and taking into account
eq. (\ref{Inversed_Sigma}), one obtains
\begin{eqnarray}
 P_{\rm heal} (t) = \dfrac{1}{2} \sqrt{\dfrac{1}{2\pi \sigma_{x x}(t)}} \:
\exp\left\lbrace  - \dfrac{x^2(t)}{2 \sigma_{x x}(t)} \right\rbrace  \left\{1 -
{\rm erf} \left[ \dfrac{\sigma_{x x}(t) \: v(t) - \sigma_{x v}(t) \:
x(t)}{\sqrt{2 \sigma_{x x}(t) \det {\bm \sigma}}}\right] \right\}
\label{Healing_PDF}
\end{eqnarray}
where the error function ${\rm erf} (z) = (2/\sqrt{\pi}) \int_{0}^{z} \:
{\rm e}^{- x^2} dx$, and $x(t)$, $v(t)$ are given by eq. (\ref{Mean_Values})
(with $x' = 0$ and $v' = |\kappa|$).

\subsubsection{Healing distance distribution function}

In order to determine the distribution of distances $h$ from the hump at $x=0$
where the extending bond may still turn back to shrinking, one may consider the
PDF for the maximal divergence distance $Q(h)$ before the bond returns and
heals. In this case a critical value $h_c$ can be defined as a point where $Q
(h)$ gets very small.

One may express $Q(h)$ in terms of the transition probability $P (x, v, t|
x', v', t')$ by taking into consideration the following arguments. The overall
return process can  be seen as the composition (recall that the process is
Markovian) of two processes. The first one starts from the top of the inverted
harmonic potential with the initial $x' = 0$  and $v' = |\kappa|$ and continues
down the ramp until a turning point $x = h$ where the velocity becomes zero,
i.e., $v = 0$, at an intermediate time moment $t'$. The probability of this
process is given by $P (h, 0, t'| 0, |\kappa|, 0)$. The reverse process starts
from the turning point (i.e., $x'=h, v' = 0$) at time moment $t'$,  and
continues back to the top of the potential where $x = 0$ and the velocity can be
any in the interval $-\infty \leq v \leq 0$. The latter means that the
corresponding transition probability should be integrated over velocity,
i.e., $\int_{-\infty}^{0} dv P (0, v, t| h, 0, t')$. Finally, since the
total time interval may also be arbitrary, one should integrate over times.
Thus, one obtains
\begin{eqnarray}
 Q(h) &=& \int\limits_{0}^{\infty} dt\int\limits_{0}^{t}
dt'\int\limits_{-\infty}^{0} dv \: P (0, v, t| h, 0, t') \:  P (h, 0, t'| 0,
|\kappa|, 0)\nonumber\\
&=& \underbrace{\int\limits_{0}^{\infty} d\tau \int\limits_{-\infty}^{0} dv \: P
(0, v, \tau| h, 0, 0)}_{I(h)} \: \underbrace{ \int\limits_{0}^{\infty} dt' P (h,
0, t'|0, |\kappa|, 0)}_{J(h)} \nonumber 
= I (h) \: J(h)
\label{Q}
\end{eqnarray}
where we have changed the order of time integration. Taking into account the
form of the transition probability, eq. (\ref{General_Solution}), and after 
integrating over the velocity, the expression for $I(h)$ reads
\begin{eqnarray}
 I(h) = \dfrac{1}{2} \: \int\limits_{0}^{\infty} dt \sqrt{\dfrac{1}{2 \pi
\sigma_{x x} (t)}} \: \exp \left\{ - \dfrac{x^2 (t)}{2 \sigma_{x x} (t)}
\right\} \left\{ 1 - {\rm erf} \left[  \dfrac{\sigma_{x x} (t) v(t) - \sigma_{x
v} (t) x (t)}{\sqrt{2 \sigma_{x x} (t) \det {\bm \sigma}}}\right] \right\}
\label{I}
\end{eqnarray}
where $x (t)$ and $v(t)$ are given by
\begin{eqnarray}
 x(t) &=& G_{x x} (t) \: h\nonumber\\
v(t) &=& G_{v x} (t) \: h
\end{eqnarray}

By making use of the expressions for the inverse $\sigma$-matrix, eq.
(\ref{Inversed_Sigma}), the expression for $J(h)$ takes on the form
\begin{eqnarray}
J(h) = \dfrac{1}{2 \pi} \: \int\limits_{0}^{\infty} dt \dfrac{1}{\sqrt{\det
{\bm \sigma}}} \: \exp \left\{ - \dfrac{\sigma_{v v} (t) [h - x(t)]^2 + 2
\sigma_{x v} (t) [h - x(t)] v(t) + \sigma_{x x} (t) v^2 (t)}{2 \det {\bm
\sigma}} \right\}
\label{J}
\end{eqnarray}
with
\begin{eqnarray}
 x(t) &=& G_{x v} (t) \: |\kappa|\nonumber\\
v(t) &=& G_{v v} (t) \: |\kappa|\;.
\end{eqnarray}

\subsection{MD-simulation of Self-Healing}

The simulation scheme for sampling the self-healing distributions is as follows
- once the largest bond in the chain crosses the barrier position, one monitors
its expansion for $1000$ integration steps and records the maximum expansion and
the time spent before it crosses back the barrier. Frequently such a 
pseudo-broken bond heals again so that from the above record, one may recover
the distribution of the maximum expansion beyond the barrier, and the time
spent in a pseudo-broken state before the healing occurs.
\begin{figure}[ht]
\vspace{0.5cm}
\includegraphics[height = 5.0 cm, width = 6.0 cm]{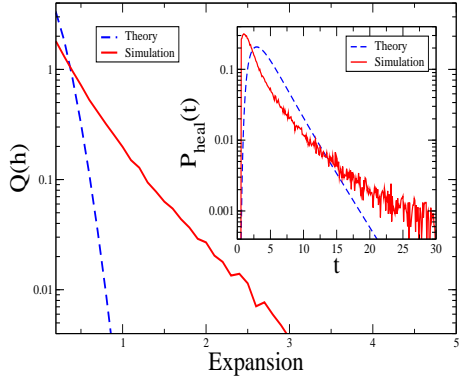}
\caption{Probability distribution of healing bond lengths and healing times
(inset) for a $30$-bead chain in $1$D and $f = 0.25$. Both plots are shown in
semilog coordinates. Dashed lines denote theoretical predictions.}
\label{Heal}
\end{figure}

In Fig. \ref{Heal} we show the distribution of the healing lengths and healing
times (inset) obtained from simulations and from theory. Evidently, the
healing time distribution $P_{\rm heal}(t)$ looks qualitatively identical for
both the theoretical treatment (which is based on eq. (\ref{Healing_PDF}))and
the simulations: there is a maximum just beyond the barrier crossing event 
for $t \neq 0$ and a fast (exponential) decay thereafter. The existence of the
maximum can be attributed to the
thermostat - just after crossing the barrier one has to wait for a while
a thermal kick turns the trajectory into reverse direction. On the
other hand, for longer time intervals the downhill motion wins (i.e.,
the healing becomes progressively improbable) and $P_{\rm heal}(t)$ decreases
rapidly with time.
\begin{figure}[ht]
\vspace{0.5cm}
\centering
\includegraphics[height = 5.0 cm, width = 6.0 cm]{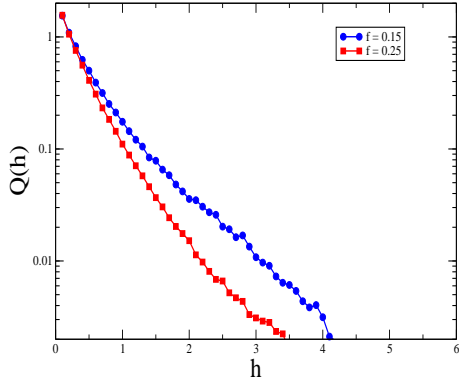}
\caption{The healing expansion distribution $Q(h)$ for $f = 0.15$ and $f =
0.25$ plotted in semi-logarithmic coordinates. The zero of the x-axis
corresponds to the barrier positions, which are different for the two different
values of the force.}
\label{Force}
\end{figure}

The distribution of the healable expansions beyond the
barrier is of greater importance because it sets a more stringent criterion
of true (irreversible) rupture. We see that for $f = 0.25$, the
probability of healing is negligible for an expansion of $\sim 2.0$ beyond the
barrier, i.e., a bond expansion of $2.92 + 2.0 \approx 5.0$. Once the expansion
of the bond reaches $h_c \approx 5.0$ for $f = 0.25$, it is highly unlikely for
the bond to heal. One should note that the process is stochastic in nature so
that a bond that has expanded even beyond $h_c$ may still manage to heal. Such
an event, however, is highly unlikely and the criterion defined in the above
manner serves to be a good one for all practical purposes. The theoretical
distribution (calculated according to eqs. (\ref{Q}), (\ref{I}) and (\ref{J})),
decays much faster than that obtained from simulations; this is because of the
faster descent of the quadratic potential (considered in theory within the
linearized approximation) as compared to the roughly linearly decaying effective
potential beyond the barrier.

     How does the above picture change when the conditions of the simulation are
varied? To answer this question, we have performed simulations and we find that
the above distribution does not vary much with friction, chain length or whether
the simulation is performed in $1$D or $3$D. The distribution however depends on
the applied tensile force as shown in Fig. \ref{Force} for two different forces.

 As expected intuitively, the distribution decays faster for a larger force
because healing is more unlikely in case of a steeper unstable potential beyond
the barrier. From a practical point of view, such a distribution for several
different forces can be very helpful in determining a criterion for irreversible
rupture.

 We next turn to the question of the dependence of the healing on temperature.

\begin{figure}[ht]
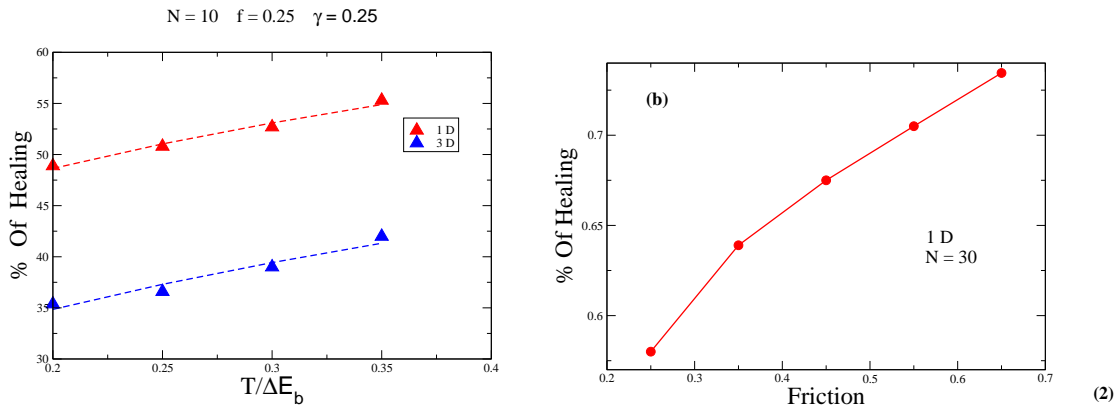

\vspace{0.5cm}
\centering
\includegraphics[scale=0.27]{fig13a.eps}
\hspace{0.5cm}
\includegraphics[scale=0.27]{fig13b.eps}
\caption{(a) The healing fraction for $f=0.25$ and $\gamma = 0.25$ at
different values of the temperature. A power-law regression gives an exponent of
$\sim 0.2$ in $1$D and $\sim 0.3$ in $3$D. (b) Increase of the fraction of
self-healing bonds with friction $\gamma$.}
\label{Heal_Percent}
\end{figure}

Figure \ref{Heal_Percent} shows the fraction of  self-healing events on the
temperature for $f = 0.25$ and $\gamma = 0.25$. As expected, healing
becomes more frequent as the temperature is increased because the bead that has
crossed the barrier now receives stronger thermal kicks capable of sending it
back over the barrier.  One can also see that for the same temperature,
healing is (understandably) more frequent in $1D$ than in $3D$. The effect of
growing friction $\gamma$ is similar, on the one hand it strengthens kicks by
the thermostat on the beads at the verge of breaking, and on the other hand it
delays such beads in ``rolling down'' the downhill ramp potential,  providing
more time to receive a thermal kick which could turn them back.

\section{Conclusion}
\label{Summary}

In the present work we have studied the process of polymer chain breakage for
linear chains subjected to constant tensile force by multidimensional
Kramers-Langer approach and compared theoretical predictions to results of
extensive MD simulations in one- and three dimensions. The adopted theoretical
treatment makes it possible to consider collective unstable modes as being
mainly responsible for chain scission. Comparison with simulation data
as, for example, the distribution of the probability for scission over
bond index, the scission time distribution, variation of MFPT with temperature
and chain length as well as the bond breaking dynamics indicates that the
Kramers-Langer approach agrees qualitatively with observations.

We demonstrate that the recombination (self-healing) dynamics of the breaking
bonds can be qualitatively reproduced by the Kramers equation. We derive
analytic expressions for the healing time and expansion distributions in
reasonable agreement with the MD data and reveal the variation of the fraction
of healing bonds with temperature and friction. In fact, we demonstrate that
more that $50\%$ of the crossings of the activation barrier end up
as healed bonds again (this fraction changes with temperature) especially in the
$3D$ simulations. This finding should be kept in mind in the assessment of
earlier research work where only $1D$ simulations of breaking chains have been
performed and bond-healing was not allowed for.

One should emphasize, however, that we still find a discrepancy of nearly an
order of magnitude between theory and computer experiment regarding the mean
rate of scission, that is, the rate of bond breaking is significantly
underestimated by theory (cf. also \cite{Sain}). We believe that the origin of
this discrepancy is related to the existence of non-linear localized excitations
(breathers) in the anharmonic lattice which remain out of the scope of this
investigation. Recent studies \cite{Peyrard} indicate the breathers play very
important role regarding energy transfer in discrete non-linear chains which
serve a generic model for biopolymers. Clearly, furter research in this
direction is needed before a good understanding of the whole problem is reached.

\section{Acknowledgments}
We are indebted to B. D\"unweg and L. Manevitch for helpful discussions in the
course of this study. This work has been supported by the Deutsche
Forschungsgemeinschaft (DFG),  Grant No. SFB 625/B4.

\begin{appendix}
 \section{Calculation of $\sum_{i, j} \nu_i \Gamma_{i j} \nu_j$}
\label{Calculation}

Let us calculate $\sigma \equiv \sum_{i, j} \nu_i \Gamma_{i j} \nu_j$ which
occurs in eq. (\ref{Lambda_Plus}). Because $\sum_{i=1}^{2N} \nu_i^2 = 1$ this
can be seen as a eigenvalue problem, i.e.
\begin{eqnarray}
 \sum_{i=1}^{2N} \: \nu_i \: \Gamma_{i j} = \sigma \nu_j
\label{Can_be_Seen}
\end{eqnarray}
The corresponding characteristic equation reads
\begin{eqnarray}
 \det [\Gamma_{i j} - \delta_{i j} \sigma] = 0
\label{GG}
\end{eqnarray}
The matrix $\Gamma_{i j}$ has a block structure (see eq. (\ref{Gamma_and_A}))
so that eq. (\ref{GG}) takes the form
\begin{eqnarray}
 \det \left(\begin{array}{ccc}
 \begin{array}{ccc}
 - \sigma &\cdots & 0\\
\vdots &   & \vdots\\
0 & \cdots & - \sigma
\end{array} &   &
\begin{array}{ccc}
 0 &\cdots & 0\\
\vdots &  \ddots & \vdots\\
0 & \cdots & 0
\end{array}
\\
 \begin{array}{ccc}
 0 &\cdots & 0\\
\vdots &  \ddots & \vdots\\
0 & \cdots & 0
\end{array}&  &\begin{array}{ccc}
 m\gamma-\sigma &\cdots & 0\\
\vdots &  \ddots  & \vdots\\
0 & \cdots   &m\gamma-\sigma
\end{array}
 \end{array}\right) = 0
\label{AA}
\end{eqnarray}

On the other hand the determinant of he block-diagonal matrix $\det
\left(\begin{smallmatrix}
       {\bf A} & {\bf 0}\\
{\bf 0} & {\bf D}
      \end{smallmatrix}\right) = \det {\bf A} \det {\bf D} = (- \sigma)^N
(m\gamma - \sigma)^N = 0$. As a result
\begin{eqnarray}
 \sigma = m \gamma
\label{Sigma}
\end{eqnarray}

\section{How to calculate the determinant of a symmetrical tridiagonal matrix}
\label{Appendix_2}

The typical tridiagonal $N\times N$-matrix which is common in the context of
one-dimensional string of beads model has the following form
\begin{eqnarray}
 {\cal T} (N) =  \left(\begin{array}{rrrrrr}
2 & -1 &  & & &  {\bf 0}\\
-1 & 2 & -1 & & &         \\
0 & -1 & 2 & -1 & &       \\
 & \ddots &\ddots  & \ddots &\ddots\\
 &  &\ 0 &-1 & 2 & -1\\
{\bf 0} &  & & &-1 &1 - \beta
 \end{array}\right)
\label{T_Matrix}
\end{eqnarray}
where $\beta$ is an arbitrary rational number. To calculate the determinant we
use the Laplace's formula \cite{Lancaster} which holds that for an arbitrary
matrix $A$ the determinant $\det ( A ) = \sum_{j=1}^N \: A_{i j} (-1)^{i+j}
M_{i j}$ where $M_{i j}$ is the minor (i.e. the determinant of the matrix that
results from A by removing the i-th row and the j-th column). By making use
this formula for the matrix in eq.(\ref{T_Matrix}) with
respect to the first row and expanding similarly again, one finds the recurrence
relation
\begin{eqnarray}
 \det[ {\cal T} (N)] = 2 \det[ {\cal T} (N-1)] - \det[ {\cal T} (N-2)]
\label{Recurrence}
\end{eqnarray}
This reccurence relation to be solved needs two initial conditions
\begin{eqnarray}
 \det[{\cal T} (1)] = 1- \beta \qquad \det[{\cal T} (2)] =
\det \left(\begin{array}{rr}
       2 & -1\\
       -1 & {1 - \beta}
      \end{array}\right)=1-2 \beta
\end{eqnarray}
Thus
\begin{eqnarray}
\begin{array}{lllllll}
&\det[{\cal T} (3)] &= &2(1- 2 \beta) &- &{1 + \beta} &= 1- 3 \beta\\
\\
&\det[{\cal T} (4)] &= &2(1- 3 \beta) &- &{1 + 2 \beta} &= 1- 4 \beta\\
\\
\multicolumn{7}{l}{\dotfill}\\
\\
&\det[{\cal T} (N)] &=  &{1- \beta N} & & &
\end{array}
\end{eqnarray}
As a result we can
finally write down
\begin{eqnarray}
 \det[{\cal T} (N)] = (1 - \beta N)
\label{Det_Final}
\end{eqnarray}

\section{Solution of one-dimensional Kramers equation}
\label{K_Equation_Solution}

The general solution of eq. (\ref{Kramers_Equation}) reads \cite{Risken}
\begin{eqnarray}
 P (x, v, t| x', v', 0) &=& \dfrac{1}{2 \pi (\det {\bm \sigma})^{1/2}} \: \exp
\Bigl\{ - \dfrac{1}{2} [{\bm \sigma}^{-1} (t)]_{x x} \: [x - x(t)]^2  - [{\bm
\sigma}^{-1} (t)]_{x v} \: [x - x(t)] \: [v - v(t)] \nonumber\\
&-& \dfrac{1}{2} [{\bm \sigma}^{-1}(t)]_{v v} \: [v - v(t)]^2 \Bigr\}
\label{General_Solution}
\end{eqnarray}
where the $2\times 2$ $\sigma$-matrix
\begin{eqnarray}
 \sigma_{i j}(t) = \left(
\begin{array}{cc}
\sigma_{x x}(t) & \sigma_{x v}(t) \\
\sigma_{x v} (t)& \sigma_{v v}(t)
\end{array}\right)
\end{eqnarray}
has the following elements
has the following elements
\begin{eqnarray}
\sigma_{x x}(t) &=& \dfrac{\gamma v_{\rm th}^2}{(\lambda_1 - \lambda_2)^2} \:
\left[ \dfrac{\lambda_1 + \lambda_2}{\lambda_1 \lambda_2} + 4
\dfrac{{\rm e}^{-(\lambda_1+\lambda_2) t} - 1}{\lambda_1 + \lambda_2} -
\dfrac{{\rm e}^{- 2 \lambda_1 t}}{\lambda_1} - \dfrac{{\rm e}^{- 2 \lambda_2
t}}{\lambda_2}\right] \nonumber\\
\sigma_{x v}(t) &=&  \dfrac{\gamma v_{\rm th}^2}{(\lambda_1 - \lambda_2)^2} \:
\left[{\rm e}^{-  \lambda_1 t} - {\rm e}^{-  \lambda_2 t}\right]^2\nonumber\\
\sigma_{v v}(t) &=&   \dfrac{\gamma v_{\rm th}^2}{(\lambda_1 - \lambda_2)^2}
\:\left\{ \lambda_1 + \lambda_2  + \dfrac{4 \lambda_1 \lambda_2}{\lambda_1 +
\lambda_2} \: [{\rm e}^{- (\lambda_1 + \lambda_2} - 1] - \lambda_1 \: {\rm e}^{-
 2 \lambda_1 t}  - \lambda_2 \: {\rm e}^{-  2 \lambda_2 t}\right\}
\end{eqnarray}
and where the eigenvalues $\lambda_1$ and $\lambda_2$ are
\begin{eqnarray}
 \lambda_1 &=& \dfrac{1}{2} (\gamma + \sqrt{\gamma^2 + 4 \Omega_2^2})\\
\lambda_2 &=& \dfrac{1}{2} (\gamma - \sqrt{\gamma^2 + 4 \Omega_2^2}) = \kappa <
0
\end{eqnarray}

We underline that the negative eigenvalue $\lambda_1$ is nothing but the
transmission factor $\kappa$ given by eq. (\ref{Kappa_Eqation}) whereas the
characteristic frequency $\Omega_2$ is given by eq. (\ref{Frequency}).
The determinant and the inverse $\sigma$-matrix in eq. (\ref{General_Solution})
are defined as as follows
\begin{eqnarray}
\det {\bm \sigma} &=& \sigma_{x x} \sigma_{v v} - \sigma_{x v}^2\nonumber\\
({\bm \sigma}^{-1})_{x x} &=& \dfrac{\sigma_{v v}}{\det {\bm \sigma}}\nonumber\\
({\bm \sigma}^{-1})_{x v} &=& ({\bm \sigma}^{-1})_{v x} = - \dfrac{\sigma_{x
v}}{\det {\bm \sigma}}\nonumber\\
({\bm \sigma}^{-1})_{v v} &=& \dfrac{\sigma_{x x}}{\det {\bm \sigma}}
\label{Inversed_Sigma}
\end{eqnarray}

The mean values $x(t)$  and $v(t)$ in eq. (\ref{General_Solution}) are given by
\begin{eqnarray}
x(t) &=& G_{x x} (t) \: x' + G_{x v} (t) \: v'\nonumber\\
v(t) &=& G_{v x} (t) \: x' + G_{v v} (t) \:  v'
\label{Mean_Values}
\end{eqnarray}
where the Green function matrix elements read
\begin{eqnarray}
  G_{x x} (t) &=& \dfrac{\lambda_1{\rm e}^{-  \lambda_2 t} - \lambda_2 {\rm
e}^{-  \lambda_1 t}}{\lambda_1 - \lambda_2}  \nonumber\\
G_{x v} (t) &=& \dfrac{{\rm e}^{-  \lambda_2 t} -  {\rm
e}^{-  \lambda_1 t}}{\lambda_1 - \lambda_2} \nonumber\\
G_{v x} (t) &=& \lambda_1  \lambda_2 \: \dfrac{{\rm e}^{-  \lambda_1 t} -  {\rm
e}^{-  \lambda_2 t}}{\lambda_1 - \lambda_2}\nonumber\\
G_{v v} (t) &=& \dfrac{\lambda_1{\rm e}^{-  \lambda_1 t} - \lambda_2 {\rm
e}^{-  \lambda_2 t}}{\lambda_1 - \lambda_2}
\end{eqnarray}

\end{appendix}

\end{document}